\input amstex
\documentstyle{amsppt}

\mag=\magstep1
\vsize=21.6truecm
\hsize=16truecm
\NoBlackBoxes
\leftheadtext{Yunbo Zeng and Wen-Xiu Ma}
\rightheadtext{Separation of variables for soliton equations}

\topmatter
 \title Separation of variables for soliton equations \\
via  their binary constrained flows
\endtitle
\author 
Yunbo Zeng* \footnote { E-mail:yzeng\@tsinghua.edu.cn} 
and Wen-Xiu Ma$^{\dagger}$\footnote {E-mail:mawx\@cityu.edu.hk}
\endauthor
\affil 
*Department of Applied Mathematics,
Tsinghua University\\ Beijing 100084, China\\
$\dagger$ Department of Mathematics,
City University of Hong Kong\\
Kowloon, Hong Kong, China\\
\endaffil
\endtopmatter 
\def\p{{\partial}}
\def\la{{\lambda}}
\def\La{{\Lambda}}

\TagsOnRight
{\bf Abstract.} Binary constrained flows of soliton equations admitting 
$2\times 2$ Lax matrices have $2N$  degrees of freedom, which is twice as
 many as degrees of freedom in the case of mono-constrained flows. For their 
separation of variables 
 only $N$ pairs of canonical separated variables can be introduced via their 
Lax matrices by using the normal method. A new method to introduce the other 
$N$ pairs of canonical separated variables and additional separated equations 
is proposed. The Jacobi inversion problems for binary constrained flows are 
established. Finally, the factorization of soliton equations by two commuting 
binary constrained flows and the separability of  binary constrained flows 
enable us to construct the Jacobi inversion problems for some soliton 
hierarchies.
\par
\ \par
{{\bf Keywords}:} binary constrained flow, separation of variables, Jacobi 
inversion problem,  Lax representation, factorization of soliton equations.\par

\ \par

\newpage

\subhead {1. Introduction}\endsubhead\par
\ \par
The separation of variables is one of the most universal methods for solving 
completely integrable (classical and quantum) models. It has been applied 
successfully to the study of a large number of finite-dimensional integrable 
Hamiltonian systems (FDIHSs) (see, for example, [1-12]), as well as infinite 
dimensional integrable Hamiltonian systems in the determination of 
finite-dimensional quasi-periodic solutions (see, for example, [13-18]).
In many cases the separation of variables of integrable classical systems 
prepares the passage to the corresponding quantum systems. For the classical 
integrable systems subject to the inverse scattering method, the standard 
construction of the action-angle variables using the poles of the 
Baker-Akhiezer function is in fact equivalent to the separation of 
variables [4].\par
For a FDIHS, let $m$ denote the number of degrees of freedom, and $P_i,
 i=1,...,m,$ be
functionally independent integrals of motion in involution, the separation 
of variables means to construct $m$ pairs of canonical separated variables 
$v_k, u_k, k=1,...,m$,[2,3,4]
$$\{u_k, u_l\}=\{v_k, v_l\}=0,\qquad \{v_k, u_l\}=\delta_{kl},
\qquad k,l=1,...,m, \tag 1.1$$
and $m$ functions $f_k$ such that
$$f_k(u_k, v_k, P_1,...,P_{m})=0, \qquad k=1,...,m.\tag 1.2$$
The equations (1.2) are called separated equations, which give rise to an 
explicit factorization of the Liouville tori.\par
For the FDIHSs with the Lax matrices admitting the $r$-matrices of the
 $XXX, XXZ$ and $XYZ$ type, there is a general approach to introduce 
canonical separated variables
[2,3,4,8]. The corresponding separated equations enable us to express 
the generating function of canonical transformation in completely separated 
form as an abelian integral on the associated invariant spectral curve. The
 resulted linearizing map is essentially the Abel map to the Jacobi variety 
of the spectral curve, thus providing a link, through purely Hamiltonian 
methods, with the algebro-geometric linearization methods given by [19-22].\par
 An important feature of the separation of variables for a FDIHS is that the 
number of canonical separated variables $u_k$ should be equal to the number 
$m$ of degrees of freedom. In some cases, the number of $u_k$ resulted by 
the normal method may be less than $m$ and so some additional canonical 
separated variables should be introduced. So far very few models in these
 cases have been studied. These cases remain to be a challenging problem [4].
In recent years binary constrained flows of soliton hierarchies have attracted
  attention (see, for example, [23-29]), whose basic idea was described 
in [30]. The binary constrained flows are a kind of FDIHSs for which the 
method presented in [2,3,4,8] is not valid. The degree of freedom  for 
binary constrained flows admitting $2\times 2$ Lax matrices is an even 
natural number usually denoted by $2N$. The method in [2,3,4,8] allows 
us to introduce only $N$ pairs of canonical separated variables $u_1,...,u_N$
 and $v_1,...,v_N$ via the Lax matrices. In this paper we propose a new 
method for determining  additional $N$ pairs of canonical separated variables 
and separated equations for binary constrained flows. The main idea is to
 construct two functions $\widetilde B(\la)$ and $\widetilde A(\la)$
 defining $u_{N+1},...,u_{2N}$ by the set of zeros of $\widetilde B(\la)$ 
and $v_{N+k}=\widetilde A(u_{N+k})$. To keep the canonical conditions (1.1)
 and the requirement for the separated equations (1.2), it is found that 
certain  commutator relations should be imposed on $\widetilde B(\la), 
\widetilde A(\la)$ and $\widetilde A(\la)$ has some link with the generating 
function of integrals of motion of binary constrained flows, which provides
 a way to construct the $\widetilde B(\la)$ and $\widetilde A(\la)$. In fact,
 we have to modify the original approach for introducing $u_1,...,u_N$ and 
$v_1,...,v_N$ so that $u_1,...,u_{2N}$ and $v_1,...,v_{2N}$ are canonical 
conjugated. Having produced the separation of variables,
we further construct the Jacobi inversion problems for binary constrained 
flows. This method is somewhat different from that for introducing canonical
 variables presented in [31] and can be applied to more binary constrained 
flows.
\par
Briefly, separation of variables can be characterized as a reduction of a 
multidimensional problem to a set of one-dimensional ones. The separation 
of variables of soliton equations in this paper contains two steps of 
separation of variables. The first step is to factorize $(1+1)-$dimensional
 soliton equations into two commuting $x-$ and $t-$FDIHSs via binary 
constrained flows, namely the $x-$ and $t-$dependences of the soliton 
equations are separated by the $x-$ and $t-$FDIHSs obtained from the $x$- 
and $t$-binary constrained flows. The second step is to produce separation
 of variables for the $x-$ and $t-$FDIHDs by our method to be proposed later
 on. Finally, combining the factorization of soliton equations with the 
Jacobi inversion problems for $x-$ and $t-$FDIHSs enables us to establish
 the Jacobi inversion problems for soliton equations. We will present the 
separation of variables for the KdV hierarchy, the AKNS hierarchy and the 
Kaup-Newell hierarchy via their binary constrained flows. In fact, we employ
 our method in a little different way for those three cases.\par 
In section 2, we recall the binary constrained flows of KdV hierarchy and 
present  factorization of the KdV equations into the $x$- and $t$-binary 
constrained flows. By means of the Lax matrix $M=\left( \matrix A(\la)&B(\la)
\\C(\la)&-A(\la)\endmatrix\right)$ for the binary constrained flows, the 
method in [2,3,4,8] allows us to define only $N$ pairs of canonical variables 
$u_1,...,u_N$  by the set of zeros of $B(\la)$ and $v_k=2A(u_k)$. We propose 
a new method to construct two new functions $\widetilde B(\la)$ and 
$\widetilde A(\la)$ for introducing $u_{N+1},...,u_{2N}$ by the set of
 zeros of $\widetilde B(\la)$ and $v_{N+k}=\widetilde A(u_{N+k})$. The 
construction of $\widetilde B(\la)$ and $\widetilde A(\la)$ is based on an 
observation that the canonical conditions (1.1) need certain commutator 
relations between $\widetilde A(\la), \widetilde B(\la)$, and the requirement 
for the separated equations (1.2) links $\widetilde A(\la)$ with another 
generating function of integrals of motion. To guarantee that $v_1,...,v_{2N}$
 and $u_1,...,u_{2N}$ are canonical conjugated, we also have to modify 
the original way for introducing $u_1,..,u_N$ and $v_1,...,v_N$. Then we
 establish the Jacobi inversion problems for the $x$- and $t$-binary 
constrained flows. Finally, these Jacobi inversion problems together with
 the factorization of the KdV equations give rise to the Jacobi inversion 
problems for the KdV equations. In section 3, the factorization of the 
AKNS equations is given. Since $B(\la)$ for the binary constrained AKNS 
flows, unlike the $B(\la)$ for the binary constrained KdV flows, has only
 $N-1$ zeros, we have to modify the method proposed in section 2 in order 
to find $2N$ pairs of canonical variables for the binary constrained AKNS 
flows. In section 4, we present the factorization of the Kaup-Newell 
equations. Since the commutator relations of
$A(\la), B(\la)$ and $C(\la)$ for the binary constrained Kaup-Newell 
flows are quite different from those for both the binary constrained KdV
 flows and the binary constrained AKNS flows, we need to further modify 
the method in sections 2 and 3 in order to find the separation of variables
 for the Kaup-Newell equations. Finally some remarks are made in section 5.\par

\par
\subhead {2. Separation of variables for the KdV equations}\endsubhead\par
In this section, we use the binary constrained flows of KdV hierarchy to 
illustrate our method of introducing canonical separated variables. Then 
we show how to produce the separation of variables for the KdV equations.
To make the paper self-contained, we first briefly describe the binary 
constrained flows of the KdV hierarchy [26]. 
\subhead {2.1 Binary constrained flows of the KdV hierarchy}\endsubhead\par
Let us start from the Schr$\ddot{\text o}$dinger equation [32]
$$\phi_{xx}+(\la+u)\phi=0$$
which can be rewritten as following spectral problem
$$\phi_x=U(u,\la)\phi,\quad
 U(u, \lambda)
=\left( \matrix 0&1\\-\la-u&0\endmatrix\right),\quad 
\phi=\binom {\phi_{1}}{\phi_{2}}.\tag 2.1$$
Its adjoint representation reads
$$V_x=[U, V]\equiv UV-VU.\tag 2.2$$
Set
$$V=\sum_{i=0}^{\infty}\left( \matrix a_i&b_i\\c_i&-a_i\endmatrix\right)
\la^{-i}.\tag 2.3$$
Equation (2.2) yields
$$a_{0}=b_{0}=0,\quad c_{0}=-1,\quad a_1=0,\quad b_{1}=1,\quad
c_{1}=-\frac 12u,$$
$$a_{2}=\frac 14u_x,\quad b_{2}=-\frac 12u,\quad c_{2}=\frac 18(u_{xx}+u^2),
...,$$
and in general
$$ b_{k+1}=Lb_{k}=-\frac 12L^{k-1}u,, \quad 
a_k=-\frac 12b_{k,x},\tag 2.4a$$
$$c_k=-\frac 12b_{k,xx}-b_{k+1}-b_{k}u,\qquad
 k=1,2,\cdots,\tag 2.4b$$
where 
$$L=-\frac 14\p^2-u+\frac 12\p^{-1}u_x,\qquad \p=\p_x,\quad \p^{-1}\p=
\p\p^{-1}=1.$$\par
Set
$$V^{(n)}(u, \la)=\sum_{i=0}^{n+1}
\left( \matrix a_i&b_i\\c_i&-a_i\endmatrix\right)\la^{n+1-i}+\left( \matrix
 0&0\\b_{n+2}&0\endmatrix\right),\tag 2.5$$
and take the time evolution law of $\phi$ as
$$\phi_{t_n}=V^{(n)}(u,\la)\phi.\tag 2.6$$
Then the compatibility condition of the equations (2.1) and (2.6) gives rise 
to the n-th KdV equation which can 
be written as the infinite-dimensional
Hamiltonian system
$$u_{t_n}=-2b_{n+2,x}=\p L^nu=\p\frac {\delta H_{n}}{\delta u}, \tag 2.7$$
where the Hamiltonian $H_n$ is given by
$$H_{n}=\frac{4b_{n+3}}{2n+3},\qquad
\frac {\delta H_{n}}{\delta u}=-2b_{n+2}.$$\par
The matrix $V$ determined by (2.2) and (2.3) also satisfies the adjoint 
representation of (2.6)
$$V_{t_n}=[V^{(n)}, V],\tag 2.8$$
when $u$ satisfies (2.7). \par For $n=1$ we have
$$\phi_{t_1}=V^{(1)}(u,\la)\phi,\qquad
V^{(1)}=\left( \matrix \frac 14u_x&\la-\frac 12u\\-\la^2-\frac 12u\la+\frac
 14u_{xx}+\frac 12u^2&-\frac 14u_x\endmatrix\right),\tag 2.9$$
and the equation (2.7) for $n=1$ is the well-known KdV equation
$$u_{t_1}=-\frac 14(u_{xxx}+6uu_x). \tag 2.10$$\par 
The adjoint spectral problem reads
$$\psi_x=-U^T(u,\la)\psi,\qquad
\psi=\binom {\psi_{1}}{\psi_{2}}.\tag 2.11$$
We have [26]
$$\frac {\delta\la}{\delta u}
=\beta Tr[\left(\matrix \phi_1\psi_1
& \phi_1\psi_2\\ \phi_2\psi_1&\phi_2\psi_2
\endmatrix\right)\frac {\p U(u,\la)}{\p u}]=-\beta\psi_2\phi_1,\tag 2.12$$
where $\beta$ is some constant.\par
The binary $x$-constrained flows of the KdV hierarchy (2.7) consist
 of the equations obtained from
 the spectral problem  (2.1) and the adjoint spectral problem (2.11) for $N$
distinct real numbers $\lambda_j$ and the restriction of the variational
derivatives for the conserved
quantities $H_{k_0}$ (for any fixed $k_0$) and $\lambda_{j}$:
$$ \Phi_{1,x}=\Phi_{2},\qquad\Phi_{2,x}=-\La\Phi_{1}-u\Phi_{1}
,\tag 2.13a$$
$$ \Psi_{1,x}=\La\Psi_{2}+u\Psi_{2},\qquad\Psi_{2,x}=-\Psi_{1},\tag 2.13b$$
$$\frac {\delta H_{k_0}}{\delta u}-\beta^{-1}
\sum_{j=1}^{N}\frac {\delta \lambda_{j}}{\delta u}
=-2b_{k_0+2}+<\Psi_2,\Phi_1>=0.\tag 2.13c$$
Such a constraint (2.13c)
 has been recognized as a symmetry constraint [25,26,30].
Hereafter we denote the inner product in $\text{\bf R}^N$ by $<.,.>$ and
$$\Phi_i=(\phi_{i1},\cdots,\phi_{iN})^{T},\quad
\Psi_i=(\psi_{i1},\cdots,\psi_{iN})^{T},\quad i=1,2,\quad  \Lambda=diag
(\lambda_1,\cdots,\lambda_N).$$\par
For $k_0=0$, we have
$$b_{2}=-\frac 12u=\frac 12<\Psi_2,\Phi_1>,\quad i.e., \quad u=-<\Psi_2,
\Phi_1>.\tag 2.14$$
By substituting (2.14) into (2.13a) and (2.13b), the first binary 
$x$-constrained flow becomes a
 finite-dimensional Hamiltonian system (FDHS) [26]
$$ \Phi_{1x}=\frac {\p F_1}{\p \Psi_1},\quad \Phi_{2x}=\frac
 {\p F_1}{\p \Psi_2},\quad
 \Psi_{1x}=-\frac {\p F_1}{\p \Phi_1},\quad
 \Psi_{2x}=-\frac {\p F_1}{\p \Phi_2},\tag 2.15$$
with the Hamiltonian
$$F_1=<\Psi_1, \Phi_2>-<\La\Psi_2, \Phi_1>
+\frac 12<\Psi_2, \Phi_1>^2.$$\par
The binary $t_n$-constrained flows of the KdV hierarchy (2.7) are defined 
by the replicas of  
(2.6) and its adjoint system for $N$
distinct real number $\lambda_j$ $$ \binom {\phi_{1j}}{\phi_{2j}}_{t_n}=
V^{(n)}(u,\la_j)\binom {\phi_{1j}}{\phi_{2j}},\quad\binom {\psi_{1j}}
{\psi_{2j}}_{t_n}=-(V^{(n)}(u,\la_j))^T
\binom {\psi_{1j}}{\psi_{2j}},\quad j=1,...,N, \tag 2.16$$
as well as the n-th KdV equation itself (2.7)
in the case  of the higher-order constraint for $k_0\ge 1$.
Under the constraint (2.14) and the $x$-FDHS (2.15), the binary
 $t_1$-constrained flow obtained from  (2.16) with $V^{(1)}$ given 
by (2.9) can also be written as a $t_1$-FDHS
$$ \Phi_{1,t_1}=\frac {\p F_2}{\p \Psi_1},\quad \Phi_{2,t_1}=\frac 
{\p F_2}{\p \Psi_2},\quad
 \Psi_{1,t_1}=-\frac {\p F_2}{\p \Phi_1},\quad
 \Psi_{2,t_1}=-\frac {\p F_2}{\p \Phi_2},\tag 2.17$$
with the Hamiltonian
$$F_2=-<\La^2\Psi_2, \Phi_1>+<\La\Psi_1, \Phi_2>
+\frac 12<\Psi_2, \Phi_1><\La \Psi_2, \Phi_1>$$
$$+\frac 12<\Psi_2, \Phi_1><\Psi_1, \Phi_2>
+\frac 18(<\Psi_2, \Phi_2>-<\Psi_1, \Phi_1>)^2.$$\par
The Lax representation for the $x$-FDHS (2.15) and the $t_1$-FDHS (2.17) 
can be deduced from the adjoint representation (2.2) and (2.8) by using 
the method in [33,34]
$$M_x=[\widetilde U, M],\qquad M_{t_n}=[\widetilde V^{(n)}, M],\tag 2.18$$
where $\widetilde U$ and $\widetilde V^{(n)}$ are obtained from $U$ and 
$V^{(n)}$ by a substitution of (2.14), and the Lax matrix $M$ is given by
$$M=\left( \matrix A(\la)&B(\la)\\C(\la)&-A(\la)\endmatrix\right),\tag 2.19$$
$$A(\la)=\frac{1}{4}\sum_{j=1}^{N}\frac{\psi_{1j}\phi_{1j}-\psi_{2j}\phi_{2j}}
{\la-\la_{j}},\qquad
B(\la)=1+\frac 12\sum_{j=1}^{N}\frac{\psi_{2j}\phi_{1j}}{\la-\la_{j}},$$
$$C(\la)=-\la+\frac 12<\Psi_2, \Phi_1>
+\frac 12\sum_{j=1}^{N}\frac{\psi_{1j}\phi_{2j}}{\la-\la_{j}}. $$
The equation (2.18) implies that $\frac 12Tr M^2(\la)=A^2(\la)+B(\la)C(\la)$
is the generating function of  integrals of motion for (2.15) and (2.17). 
A straightforward calculation yields
$$A^2(\la)+B(\la)C(\la)\equiv P(\la)=-\la+
\sum_{j=1}^{N}[\frac{P_{j}}{\la-\la_{j}}+\frac{P^2_{N+j}}{(\la-\la_{j})^2}], 
\tag 2.20$$
where $P_j, j=1,...,2N$, are $2N$ independent integrals of motion for the 
FDHSs (2.15) and (2.17)
$$P_j=\frac 12\psi_{1j}\phi_{2j}+(-\frac 12\la_j+\frac 14<\Psi_2, \Phi_1>)
\psi_{2j}\phi_{1j}$$
$$+\frac 18\sum_{k\neq j}\frac{1}{\la_j-\la_{k}}[(\psi_{1j}\phi_{1j}-
\psi_{2j}\phi_{2j})(\psi_{1k}\phi_{1k}
-\psi_{2k}\phi_{2k})+4\psi_{1j}\phi_{2j}\psi_{2k}\phi_{1k}],\qquad j=1,...,N
\tag 2.21a$$
$$P_{N+j}=\frac 14(\psi_{1j}\phi_{1j}+\psi_{2j}\phi_{2j}),\qquad j=1,...,N.
\tag 2.21b$$
It is easy to verify that
$$F_1=2\sum_{j=1}^{N}P_{j}, \qquad
F_2=2\sum_{j=1}^{N}(\la_jP_{j}+P^2_{N+j}). \tag 2.22$$\par
With respect to the standard Poisson bracket it is found that
$$\{A(\la), A(\mu)\}=\{B(\la), B(\mu)\}=\{C(\la), C(\mu)\}=0,\tag 2.23a$$
$$\{A(\la), B(\mu)\}=\frac 1{2(\la-\mu)}[B(\mu)-B(\la)],\tag 2.23b$$
$$\{A(\la), C(\mu)\}=\frac 1{2(\la-\mu)}[C(\la)-C(\mu)],\tag 2.23c$$
$$\{B(\la), C(\mu)\}=\frac 1{\la-\mu}[A(\mu)-A(\la)].\tag 2.23d$$
It follows from (2.23) that
$$\{A^2(\la)+B(\la)C(\la),  A^2(\mu)+B(\mu)C(\mu)\}=0,$$
which implies that $P_j, j=1,...,2N,$ are in involution:
$$\{P_k, P_l\}=0,\qquad k,l=1,...,2N.$$
Therefore the FDHSs (2.15) and (2.17) are integrable and commute with each
 other.
The construction of (2.15) and (2.17) ensures that if $(\Psi_1, \Psi_2, \Phi_1,
\Phi_2)$ satisfies the finite-dimensional integrable Hamiltonian systems 
(FDIHSs) (2.15) and (2.17) simultaneously, then $u$ defined by (2.14) solves
 the KdV equation (2.10).\par
In general, by substituting (2.14) and using (2.15), the $t_n$-constrained 
flow (2.16) becomes a $t_n$-FDIHS and the n-th KdV equation (2.7) is 
factorized by the $x$-FDIHS (2.15) and the $t_n$-FDIHS. Set
$$A^2(\la)+B(\la)C(\la)=
\la\sum_{k=0}^{\infty}\widetilde F_k\la^{-k}, \tag 2.24a$$
where $\widetilde F_k, k=1,2,...,$ are also integrals of motion for both 
the $x$-FDHSs (2.15) and the $t_n$-binary constrained flows (2.16). 
Comparing (2.24a) with (2.20), one gets
$$\widetilde F_0=-1,\quad \widetilde F_1=0, \quad \widetilde F_k=
\sum_{j=1}^{N}[\la_j^{k-2}P_j+(k-2)\la_j^{k-3}P_{N+j}^2],\quad k=2,3,.... 
\tag 2.24b$$
By employing the method in [34,35], the $t_n$-FDIHS obtained from the 
$t_n$-constrained flow (2.16) is found to be of the form
$$ \Phi_{1,t_n}=\frac {\p F_{n+1}}{\p \Psi_1},\quad \Phi_{2,t_n}=\frac 
{\p F_{n+1}}{\p \Psi_2},\quad
 \Psi_{1,t_n}=-\frac {\p F_{n+1}}{\p \Phi_1},\quad
 \Psi_{2,t_n}=-\frac {\p F_{n+1}}{\p \Phi_2},\tag 2.25a$$
with the Hamiltonian
$$F_{n+1}=\sum_{m=0}^{n}(\frac 12)^{m-1}\frac{\alpha_m}{m+1}\sum_{l_1+...
+l_{m+1}=n+2}\widetilde F_{l_1}...\widetilde F_{l_{m+1}}, \tag 2.25b$$
where $l_1\geq 1,...,l_{m+1}\geq 1, \alpha_0=1, \alpha_1=\frac 12, \alpha_2
=\frac 32,$ and [34,35]
$$\alpha_m=2\alpha_{m-1}+\sum_{l=1}^{m-2}\alpha_{l}\alpha_{m-l-1}-\frac 12
\sum_{l=1}^{m-1}\alpha_{l}\alpha_{m-l}, \quad m\geq 3.\tag 2.25c $$
The n-th KdV equation (2.7) is factorized by the $x$-FDIHS (2.15) and the 
$t_n$-FDIHS (2.25).\par
For example, for the second equation in the KdV hierarchy (2.7) with $n=2$
$$u_{t_2}=\frac 1{16}(u_{xxxx}+10uu_{xx}+5u_x^2+10u^3)_x, \tag 2.26$$
the Hamiltonian $F_3$ for the $t_2$-FDIHS reads
$$F_3=2\widetilde F_4+\frac 12\widetilde F_2^2=2\sum_{j=1}^{N}(\la_j^2P_{j}
+2\la_jP^2_{N+j})+\frac 12(\sum_{j=1}^{N}P_{j})^2. \tag 2.27$$
Then the second KdV equation (2.26) is factorized by the $x$-FDIHS (2.15) 
and the $t_2$-FDIHS with the Hamiltonian $F_3$.\par  
\ \par
\subhead {2.2 The separation of variables for the KdV equations}\endsubhead\par
\ \par
An effective way to introduce the separated variables $v_k, u_k$ and to 
obtain the separated equations  is to use the Lax matrix $M$ and the
 generating function of integrals of motion.  For the FDIHSs (2.15) 
and (2.17), we can define the first $N$ pairs of the canonical variables 
$u_k, v_k, k=1,...,N$, by the method [2,3,4,8]. The commutator relations 
(2.23) and the generating function of integrals of motion (2.20) enable us to 
define $u_1,...,u_N$ by the set of zeros of $B(\la)$
$$B(\la)=1+\frac 12\sum_{j=1}^{N}\frac{\psi_{2j}\phi_{1j}}
{\la-\la_{j}}=\frac {R(\la)}{K(\la)},\tag 2.28a$$
where
$$R(\la)=\prod_{k=1}^{N}(\la-u_{k}), \qquad
K(\la)=\prod_{k=1}^{N}(\la-\la_{k}), $$
and $v_1,...,v_N$ by
$$v_k=A_1(u_k), \qquad k=1,...,N,\tag 2.28b$$
where
$$ A_1(\la)=2A(\la)=\frac 12\sum_{j=1}^{N}\frac{\psi_{1j}\phi_{1j}-\psi_{2j}
\phi_{2j}}
{\la-\la_{j}}.$$
As we will see later, the commutator relations (2.23) guarantee that
 $u_1,...,u_N$ and $v_1,...,v_N$ satisfy the canonical conditions (1.1).
 Then substituting $u_k$ into (2.20) gives rise to the separated equations 
$$v_k=A_1(u_k)=2\sqrt {P(u_k)}
=2\sqrt {-u_k+\sum_{j=1}^{N}[\frac{P_{j}}{u_k-\la_{j}}+\frac{P^2_{N+j}}
{(u_k-\la_{j})^2}]}
,\quad k=1,...,N.$$
Now the reason for taking our choice of $B(\la)$ and $A_1(\la)$ becomes 
apparent. \par
The FDIHSs (2.15) and (2.17) have $2N$ degrees of freedom, therefore we 
need to introduce the other $N$ pairs of canonical variables $v_k, u_k, 
k=N+1,...,2N.$ The main idea is to construct two suitable functions 
$\widetilde B(\la), \widetilde A(\la)$ in order to define $u_{N+1},...,u_{2N}$
 by the set of zeros of $\widetilde B(\la)$ and $v_{N+1},...,v_{2N}$ by 
$v_{N+k}=\widetilde A(u_{N+k})$.
The above way for introducing $u_k, v_k, k=1,...,N$ stimulates us to impose
 two requirements on $\widetilde B(\la)$ and $\widetilde A(\la)$ in order 
to construct them. First, the canonical conditions (1.1) require that
$\widetilde B(\la)$ and $\widetilde A(\la)$ satisfy
$$\{\widetilde B(\la), \widetilde B(\mu)\}=
\{\widetilde B(\la), B(\mu)\}=
\{\widetilde A(\la), \widetilde A(\mu)\}=
\{\widetilde A(\la), B(\mu)\}=
\{\widetilde A(\la), A_1(\mu)\}=0,\tag 2.29a$$
$$\{\widetilde A(\la), \widetilde B(\mu)\}=\frac 1{\la-\mu}[\widetilde
 B(\mu)-\widetilde B(\la)],\tag 2.29b$$
$$\{A_1(\la), \widetilde B(\mu)\}=0.\tag 2.29c$$
The second requirement is that the equation $v_{N+k}=\widetilde A(u_{N+k})$ 
should give rise to the separated equations. Notice that $P_{N+j}$ given by 
(2.21b) are integrals of motion for the FDIHSs (2.15) and (2.17), we 
can construct another generating function $\widetilde A(\la)$ of integrals 
of motion  by
$$\widetilde A(\la)=\frac{1}{2}\sum_{j=1}^{N}\frac{\psi_{1j}\phi_{1j}+
\psi_{2j}\phi_{2j}}
{\la-\la_{j}}=2\sum_{j=1}^{N}\frac{P_{N+j}}{\la-\la_{j}}.\tag 2.30a$$
We may use $\widetilde A(\la)$ to define 
$v_{N+1},...,v_{2N}$ since substituting $u_{N+k}$ into the equation (2.30a)
 immediately leads to the separated equations for $v_{N+k}$ and $u_{N+k}$.
It is easy to see that $\{\widetilde A(\la), B(\mu)\}=\{\widetilde A(\la), 
\widetilde A(\mu)\}=\{\widetilde A(\la), A_1(\mu)\}=0$. 
We look for $\widetilde B(\la)$ in the form
$$\widetilde B(\la)=1+\frac{1}{2}\sum_{j=1}^{N}\frac 1
{\la-\la_{j}}(\delta_1\phi^2_{1j}+\delta_2\phi_{1j}\phi_{2j}+\delta_3
\phi^2_{2j}).$$
By requiring $\widetilde B(\la)$ to satisfy (2.29a) and (2.29b), one 
gets $\delta_1=1, \delta_2=\delta_3=0,$ i.e.
$$\widetilde B(\la)=1+\frac{1}{2}\sum_{j=1}^{N}\frac{\phi_{1j}^2}
{\la-\la_{j}}.\tag 2.30b$$
But $\widetilde B(\la)$ doesn't fit (2.29c). In fact,  we have 
$$\{A_1(\la), \widetilde B(\mu)\}=\frac 1{\la-\mu}[\widetilde B(\mu)
-\widetilde B(\la)].\tag 2.31$$
However, (2.29a), (2.29b) and (2.31) enable us to replace $A_1(\la)$ by
 $\overline A(\la)$
$$\overline A(\la)\equiv A_1(\la)-\widetilde A(\la)=
-\sum_{j=1}^{N}\frac{\psi_{2j}\phi_{2j}}
{\la-\la_{j}},\tag 2.32$$
and we will redefine $v_k$ by $v_k=\overline A(u_k)$.\par
Then a straightforward calculation shows that $\overline B(\la)=B(\la), 
\overline A(\la), \widetilde B(\la), \widetilde A(\la)$ satisfy the 
following required commutator relations
$$\{\overline B(\la), \overline B(\mu)\}=
\{\widetilde B(\la), \widetilde B(\mu)\}=
\{\overline A(\la), \overline A(\mu)\}=
\{\widetilde A(\la), \widetilde A(\mu)\}=0, \tag 2.33a$$
$$\{\overline B(\la), \widetilde B(\mu)\}=
\{\overline B(\la), \widetilde A(\mu)\}=
\{\widetilde B(\la), \overline A(\mu)\}=
\{\overline A(\la), \widetilde A(\mu)\}=0,\tag 2.33b$$
$$\{\overline A(\la), \overline B(\mu)\}=\frac 1{\la-\mu}[\overline B(\mu)
-\overline B(\la)],\quad
\{\widetilde A(\la), \widetilde B(\mu)\}
=\frac 1{\la-\mu}[\widetilde B(\mu)-\widetilde B(\la)].\tag 2.33c$$
We have the following proposition.

\proclaim {Proposition 1}  Assume that $\la_j, \phi_{ij}, \psi_{ij} 
\in\text {\bf R}, i=1,2, j=1,...,N$.
Introduce the separated variables
$u_{1},...,u_{2N}$ by the set of zeros of $\overline B(\la)$ and
 $\widetilde B(\la)$:
$$\overline B(\la)=B(\la)
1+\frac 12\sum_{j=1}^{N}\frac{\psi_{2j}\phi_{1j}}
{\la-\la_{j}}=\frac {R(\la)}{K(\la)},\tag 2.34a$$
$$\widetilde B(\la)=1+\frac{1}{2}\sum_{j=1}^{N}\frac{\phi_{1j}^2}
{\la-\la_{j}}=\frac {\overline R(\la)}{K(\la)},\tag 2.34b$$
with
$$R(\la)=\prod_{k=1}^{N}(\la-u_{k}), \qquad \overline R(\la)=
\prod_{k=1}^{N}(\la-u_{N+k}),$$
and  $v_{1},...,v_{2N}$ by
$$v_{k}=\overline A(u_k)=A_1(u_k)-\widetilde A(u_k)=
-\sum_{j=1}^{N}\frac{\psi_{2j}\phi_{2j}}
{u_k-\la_{j}}, \quad k=1,...,N,\tag 2.34c$$
$$v_{N+k}=\widetilde A(u_{N+k})=
\frac{1}{2}\sum_{j=1}^{N}\frac{\psi_{1j}\phi_{1j}+\psi_{2j}\phi_{2j}}
{u_{N+k}-\la_{j}}, \quad k=1,...,N.\tag 2.34d$$
If $u_1,...,u_N,$ are single zeros of $\overline B(\la)$, then
$v_1,...,v_{2N}$ and $u_1,...,u_{2N}$ are canonically conjugated, i.e.,
 they satisfy (1.1).\par
\endproclaim
{\it{Proof.}} Notice that
$$\lim_{\la\to\la_j-0}\widetilde B(\la)=-\infty , \qquad
\lim_{\la\to\la_j+0}\widetilde B(\la)=\infty, $$
it is easy to see that
$$u_{N+1}<\la_1<u_{N+2}<\la_2<...<u_{2N}<\la_N. \tag 2.35$$
We have
$\overline B'(u_k)\neq 0, \widetilde B'(u_{N+k})\neq 0$. Hereafter the
 prime denotes the differentiation with respect to $\la$.
In what follows, we take $k,l=1,...,N$. It follows from (2.33b) that
$$0=\{u_k, \widetilde B(u_{N+l})\}=
\widetilde B'(u_{N+l})\{u_k, u_{N+l}\}+
\{u_k, \widetilde B(\mu)\}|_{\mu=u_{N+l}},$$
$$0=\{\overline B(u_{k}), u_{N+l}\}=
\overline B'(u_{k})\{u_k, u_{N+l}\}+
\{\overline B(\la), u_{N+l}\}|_{\la=u_{k}},$$
$$0=\{\overline B(u_{k}), \widetilde B(u_{N+l})\}=
\overline B'(u_k)\widetilde B'(u_{N+l})\{u_k, u_{N+l}\}+\overline B'(u_k)
\{u_k, \widetilde B(\mu)\}|_{\mu=u_{N+l}}$$
$$+\widetilde B'(u_{N+l})\{\overline B(\la), u_{N+l}\}|_{\la=u_k}+
\{\overline B(\la), \widetilde B(\mu)\}|_{\la=u_k, \mu=u_{N+l}}$$
$$=\overline B'(u_k)\widetilde B'(u_{N+l})\{u_k, u_{N+l}\}+\overline B'(u_k)
\{u_k, \widetilde B(\mu)\}|_{\mu=u_{N+l}}
$$ $$+\widetilde B'(u_{N+l})\{\overline B(\la), u_{N+l}\}|_{\la=u_k},$$
which together lead to
$\{u_k, u_{N+l}\}=0$. Similarly, $\{u_k, u_{l}\}=0, \{u_{N+k}, u_{N+l}\}=0$.
\par
Using (2.33b), (2.33c) and above results, one gets
$$\{v_k, \overline B(\mu)\}\mid_{\mu=u_l}=\{\overline A(u_k), \overline 
B(\mu)\}\mid_{\mu= u_l}$$
$$=\overline A'(u_k)\{u_k, \overline B(\mu)\}\mid_{\mu=u_l}+
[\{\overline A(\la), \overline B(\mu)\}\mid_{\la=u_k}]\mid_{\mu=u_l},$$
$$=\overline A'(u_k)[\{u_k, \overline B(u_l)\}-\overline B'(u_l)\{u_k, u_l\}]+
[\{\overline A(\la), \overline B(\mu)\}\mid_{\la=u_k}]\mid_{\mu=u_l}$$
$$=[\{\overline A(\la), \overline B(\mu)\}\mid_{\la=u_k}]\mid_{\mu=u_l}=\frac 
{\overline B(\mu)-\overline B(u_k)}{u_k-\mu}\mid_{\mu=u_l}
=-\delta_{kl}\overline B'(u_k),$$
and
$$0=\{v_k, \overline B(u_l)\}=\overline B'(u_l)\{v_k, u_l\}+
\{v_k, \overline B(\mu)\}\mid_{\mu=u_l},$$
then 
$$\{v_k, u_l\}=-\frac 1{\overline B'(u_l)}\{v_k, \overline B(\mu)\}
\mid_{\mu=u_l}
=\delta_{kl}\frac {\overline B'(u_k)}{\overline B'(u_l)}=\delta_{kl}.$$
In the same way, one gets $\{v_{N+k}, u_{N+l}\}=\delta_{kl}.$ The following 
equalities
$$\{v_k, u_{N+l}\}=\{\overline A(u_k), u_{N+l}\}=\{\overline A(\la), 
u_{N+l}\}\mid_{\la=u_k},$$
$$0=\{\overline A(\la), \widetilde B(u_{N+l})\}=\widetilde B'(u_{N+l})
\{\overline A(\la), u_{N+l}\},$$
yield $\{v_k, u_{N+l}\}=0$ and similarly $\{v_{N+k}, u_{l}\}=0$.\par
Finally,
$$\{v_k, v_{N+l}\}=\{\overline A(u_k),\widetilde A(u_{N+l})\}$$
$$=\overline A'(u_k)\{u_k, \widetilde A(\mu)\}|_{\mu=u_{N+l}}+
\overline A'(u_{N+l})\{\overline A(\la), u_{N+l}\}|_{\la=u_{k}}$$
$$=\overline A'(u_k)[\{u_k, v_{N+l}\}-\widetilde A'(u_{N+l})
\{u_k, u_{N+l}\}]$$ 
$$+\widetilde A'(u_{N+l})[\{v_k, u_{N+l}\}-\overline A'(u_{k})\{u_k, 
u_{N+l}\}]=0,$$
similarly
$$\{v_k, v_{l}\}
=\overline A'(u_k)\{u_k, v_{l}\}+\widetilde A'(u_{l})\{v_k, u_{l}\}=
-\overline A'(u_{k})\delta_{kl}
+\overline A'(u_l)\delta_{kl}=0,$$
and  $\{v_{N+k}, v_{N+l}\}=0.$
This completes the proof. \par
It follows from (2.34a) and (2.34b) that
$$ \psi_{2j}\phi_{1j}=2\frac{R(\la_j)}{K'(\la_{j})},\qquad
 \phi_{1j}^2=2\frac{\overline R(\la_j)}{K'(\la_{j})},\qquad j=1,...,N,$$
or
$$ \phi_{1j}=\sqrt{\frac{2\overline R(\la_j)}{K'(\la_{j})}},\qquad
\psi_{2j}=\frac {\sqrt{2}R(\la_j)}{\sqrt{\overline R(\la_j)K'(\la_{j})}},
\qquad j=1,...,N.\tag 2.36$$
Also (2.34a) results
$$u=-<\Psi_{2}, \Phi_{1}>=2\sum_{j=1}^{N}(u_j-\la_j).\tag 2.37$$\par
We now present the separated equations. By substituting $u_k$ into (2.20),
 $u_{N+k}$ into (2.30a) and using (2.34), one gets the separated equations
$$v_k=A_1(u_k)-\widetilde A(u_k)=2\sqrt {P(u_k)}-\widetilde A(u_k)$$
$$=2\sqrt {-u_k+\sum_{j=1}^{N}[\frac{P_{j}}{u_k-\la_{j}}+\frac{P^2_{N+j}}
{(u_k-\la_{j})^2}]}
  -2\sum_{j=1}^{N}\frac{P_{N+j}}{u_k-\la_{j}},\quad k=1,...,N, \tag 2.38a$$
$$v_{N+k}=\widetilde A(u_{N+k})
=2\sum_{j=1}^{N}\frac{P_{N+j}}{u_{N+k}-\la_{j}}, 
\qquad k=1,...,N. \tag 2.38b$$
 Replacing $v_k$ by the partial derivative $\frac {\p S}{\p u_k}$ of the 
generating function $S$ of the canonical transformation and interpreting the $P
_i$ as integration constants, 
the equations (2.38) give rise to the Hamilton-Jacobi equations which are
 completely separable and may be integrated to give the completely 
separated solution
$$S(u_1,...,u_{2N})=\sum_{k=1}^{N}[\int^{u_k}(2\sqrt {P(\la)}-\widetilde
 A(\la))d\la
+\int^{u_{N+k}}\widetilde A(\la)d\la]$$
$$=2\sum_{k=1}^{N}[\int^{u_k}\sqrt {P(\la)}d\la-
\sum_{i=1}^{N}P_{N+i}ln\mid\frac{u_k-\la_i}{u_{N+k}-\la_i}\mid]. 
\tag 2.39$$\par
The linearizing coordinates are then
$$Q_i=\frac {\p S}{\p P_i}=\sum_{k=1}^{N}\int^{u_k}\frac {1}{(\la-\la_i)\sqrt
 { P(\la)}}d\la, \quad i=1,...,N, \tag 2.40a$$
$$Q_{N+i}=\frac {\p S}{\p P_{N+i}}
=2\sum_{k=1}^{N}[\int^{u_k}\frac {P_{N+i}}{(\la-\la_i)^2\sqrt { P(\la)}}d\la
-ln\mid\frac{u_k-\la_i}{u_{N+k}-\la_i}\mid], \quad i=1,...,N. \tag 2.40b$$\par
By using (2.22), the linear flow induced by (2.15) is then given by 
$$Q_i=\gamma_i+x\frac {\p F_1}{\p P_i}=\gamma_i+2x,\quad
Q_{N+i}=2\gamma_{N+i}+x\frac {\p F_1}{\p P_{N+i}}=2\gamma_{N+i},\quad i=1,...,
N. \tag 2.41$$
Hereafter $\gamma_i, i=1,...,2N,$ are arbitrary constants. Combining the 
equation (2.40) with the equation (2.41) leads to the Jacobi inversion 
problem for the FDIHS (2.15)
$$\sum_{k=1}^{N}\int^{u_k}\frac {1}{(\la-\la_i)\sqrt {P(\la)}}d\la
=\gamma_{i}+2x,
\quad i=1,...,N, \tag 2.42a$$
$$\sum_{k=1}^{N}[\int^{u_k}\frac {P_{N+i}}{(\la-\la_i)^2\sqrt { P(\la)}}d\la
-ln\mid\frac {u_k-\la_i}{u_{N+k}-\la_i}\mid]
=\gamma_{N+i},\quad i=1,...,N. \tag 2.42b$$
The $\phi_{1j}, \psi_{2j}$ and $<\Psi_{2}, \Phi_{1}>$ defined by (2.36) and 
(2.37) are the symmetric functions of $u_k, k=1,...,2N$. If, by using the
 Jacobi inversion technique [19], 
$\phi_{1j}, \psi_{2j}$ and $<\Psi_{2}, \Phi_{1}>$  can be obtained from 
(2.42), then $\phi_{2j}, \psi_{1j}$ can be found from the first and the
 last equation in (2.15) by an algebraic calculation, respectively. The
 $(\phi_{1j}, \phi_{2j}, \psi_{1j}, \psi_{2j})$ provides the solution to 
the FDIHS (2.15). \par
By using (2.22), the linear flow induced by (2.17) is then given by 
$$Q_i=\bar\gamma_i+\frac {\p F_2}{\p P_i}t_1=\bar\gamma_i+2\la_it_1,\quad $$
$$Q_{N+i}=2\bar\gamma_{N+i}+\frac {\p F_2}{\p P_{N+i}}t_1
=2\bar\gamma_{N+i}+4P_{N+i}t_1,
\quad i=1,...,N, \tag 2.43$$
where $\bar\gamma_i$ are arbitrary constants.
Combining the equation (2.40) with the equation (2.43) leads to the Jacobi 
inversion problem for the FDIHS (2.17)
$$\sum_{k=1}^{N}\int^{u_k}\frac {1}{(\la-\la_i)\sqrt { P(\la)}}d\la
=\bar\gamma_i+2\la_it_1,\quad i=1,...,N, \tag 2.44a$$
$$\sum_{k=1}^{N}[\int^{u_k}\frac {P_{N+i}}{(\la-\la_i)^2\sqrt { P(\la)}}d\la
-ln\mid\frac {u_k-\la_i}{u_{N+k}-\la_i}\mid]
=\bar\gamma_{N+i}+2P_{N+i}t_1,\quad i=1,...,N. \tag 2.44b$$\par
Finally, since the KdV equation (2.10) is factorized by the FDIHSs (2.15) 
and (2.17), combining the equation (2.42) with the equation (2.44)  and 
using (2.37) give rise to the 
 Jacobi inversion problem for the KdV equation (2.10) 
$$\sum_{k=1}^{N}\int^{u_k}\frac {1}{(\la-\la_i)\sqrt {P(\la)}}d\la
=\gamma_{i}+2x
+2\la_it_1,\quad i=1,...,N, \tag 2.45a$$
$$\sum_{k=1}^{N}[\int^{u_k}\frac {P_{N+i}}{(\la-\la_i)^2\sqrt { P(\la)}}d\la
-ln\mid\frac {u_k-\la_i}{u_{N+k}-\la_i}\mid]
=\gamma_{N+i}+2P_{N+i}t_1,\quad i=1,...,N. \tag 2.45b$$
Notice that $u$ defined by (2.37) is the symmetric function of 
$u_k, k=1,...,N$. If, by using the  Jacobi inversion technique [19], $u$
 can be found in terms of Riemann theta functions by solving  (2.45), 
then $u$ provides the solution of the KdV equation (2.10).\par
In general, since the n-th KdV equation (2.7) is factorized by the $x$-FDIHS 
(2.15) and the $t_n$-FDIHS (2.25), the above procedure can be applied to 
find the Jacobi inversion problem for the n-th KdV equation (2.7). We have the
following proposition.
\proclaim {Proposition 2} The Jacobi inversion problem for the n-th KdV 
equation (2.7) is given by
$$\sum_{k=1}^{N}\int^{u_k}\frac {1}{(\la-\la_i)\sqrt {P(\la)}}d\la
=\gamma_{i}+2x$$
$$+t_n\sum_{m=0}^{n}(\frac 12)^{m-1}\alpha_m\sum_{l_1+...+l_{m+1}=n+2}
\la_i^{l_{m+1}-2}\widetilde F_{l_1}...\widetilde F_{l_{m}}, 
\quad i=1,...,N, \tag 2.46a$$
$$\sum_{k=1}^{N}[\int^{u_k}\frac {P_{N+i}}{(\la-\la_i)^2\sqrt { P(\la)}}d\la
-ln\mid\frac {u_k-\la_i}{u_{N+k}-\la_i}\mid]
=\gamma_{N+i}$$
$$+t_n\sum_{m=0}^{n}(\frac 12)^{m-2}\alpha_m\sum_{l_1+...+l_{m+1}=n+2}
(l_{m+1}-2)\la_i^{l_{m+1}-3}P_{N+i}\widetilde F_{l_1}...\widetilde F_{l_{m}},
 $$
$$\qquad\qquad\quad i=1,...,N, \tag 2.46b$$
where $l_1\geq 1,...,l_{m+1}\geq 1$ and $\widetilde F_{l_1},...\widetilde 
F_{l_{m}},$ are given by (2.24b). 
\endproclaim\par
For example, by using (2.27), the Jacobi inversion problem for the second 
KdV equation (2.26) is given by
$$\sum_{k=1}^{N}\int^{u_k}\frac {1}{(\la-\la_i)\sqrt {P(\la)}}d\la
=\gamma_{i}+2x+(2\la_i^2+\sum_{j=1}^{N}P_{j})t_2,\quad i=1,...,N, \tag 2.47a$$
$$\sum_{k=1}^{N}[\int^{u_k}\frac {P_{N+i}}{(\la-\la_i)^2\sqrt { P(\la)}}d\la
-ln\frac {u_k-\la_i}{u_{N+k}-\la_i}]
=\gamma_{N+i}+4\la_iP_{N+i}t_2,\quad i=1,...,N. \tag 2.47b$$
The $u$ solved from the Jacobi inversion problem (2.47) provides the solution
 for the second KdV equation (2.26).\par
 The Jacobi inversion problem for the KdV hierarchy in our case is somewhat 
different from that derived by means of the stationary equations of the 
KdV hierarchy [36], since there is an additional term
$-ln|u_k-\la_i|+ln|u_{N+k}-\la_i|$ in (2.46b).

\ \par
\subhead {3. The separation of variables for the AKNS equations}\endsubhead\par

\subhead {3.1 Binary constrained flows of the AKNS hierarchy}\endsubhead\par
\ \par
For the AKNS spectral problem [37]
$$\phi_x=U(u,\la)\phi,\quad
 U(u, \lambda)
=\left( \matrix -\la&q\\r&\la\endmatrix\right),\quad 
\phi=\binom {\phi_{1}}{\phi_{2}},\quad u=\binom {q}{r},\tag 3.1$$
its adjoint representation (2.2) and (2.3) yield
$$a_{0}=-1,\quad b_{0}=c_{0}=0,\quad a_1=0,\quad b_{1}=q,\quad
c_{1}=r,\quad a_2=\frac{1}{2}qr,...,$$
and in general
$$\binom{c_{k+1}}{ b_{k+1
}}=L \binom{c_{k}}{ b_{k}}, \qquad 
a_k=\p^{-1}(qc_k-rb_k),\qquad
 k=1,2,\cdots,\tag 3.2$$
$$L=\frac 12\left( \matrix \p-2r\p^{-1}q&2r\p^{-1}r\\-2q\p^{-1}q&-\p+2q
\p^{-1}r\endmatrix\right).$$\par
Take
$$\phi_{t_n}=V^{(n)}(u,\la)\phi.\qquad V^{(n)}(u, \la)=\sum_{i=0}^{n}
\left( \matrix a_i&b_i\\c_i&-a_i\endmatrix\right)\la^{n-i}.\tag 3.3$$
Then the compatibility condition of equations (3.1) and (3.3) gives rise 
to the AKNS hierarchy 
$$u_{t_n}={\binom {q}{r}}_{t_n}
=JL^n\binom {r}{q}
=J\frac {\delta H_{n+1}}{\delta u},\qquad n=1,2,\hdots, \tag 3.4$$
where the Hamiltonian $H_n$ and the Hamiltonian operator $J$ are given by
$$J=\left( \matrix 0&-2\\2&0\endmatrix\right),\qquad
H_{n}=\frac{2a_{n+1}}{n+1},\qquad
\binom {c_n}{b_n}
=\frac {\delta H_{n}}{\delta u},\quad n=1,2,\hdots.$$\par
For $n=2$ we have
$$\phi_{t_2}=V^{(2)}(u,\la)\phi,\qquad
V^{(2)}=\left( \matrix -\la^2+\frac 12qr&q\la-\frac 12q_x\\r\la+\frac 12r_x&
\la^2-\frac 12qr\endmatrix\right),\tag 3.5$$
and the AKNS equation (3.4) for $n=2$ reads
$$q_{t_2}=-\frac 12q_{xx}+q^2r, \qquad r_{t_2}=\frac 12r_{xx}-r^2q. 
\tag 3.6$$\par 
The adjoint AKNS spectral problem is the equation (2.11) with $U$ given by 
(3.1).
We have [25]
$$\frac {\delta\la}{\delta u}=\binom{\frac {\delta\la}{\delta q}}
{\frac {\delta\la}{\delta r}}=\beta Tr[\left(\matrix \phi_1\psi_1
& \phi_1\psi_2\\ \phi_2\psi_1&\phi_2\psi_2
\endmatrix\right)\frac {\p U(u,\la)}{\p u}]=\beta\binom {\psi_1\phi_2}{\psi_2
\phi_1}.\tag 3.7$$
\par
The binary $x$-constrained flows of the AKNS hierarchy (3.4) are defined by 
[25,29]
$$ \Phi_{1,x}=-\La\Phi_{1}+q\Phi_{2},\qquad\Phi_{2,x}=r\Phi_{1}+\La\Phi_{2}
,\tag 3.8a$$
$$ \Psi_{1,x}=\La\Psi_{1}-r\Psi_{2},\qquad\Psi_{2,x}=-q\Psi_{1}-\La\Psi_{2},
\tag 3.8b$$
$$\frac {\delta H_{k_0}}{\delta u}-
\beta^{-1}\sum_{j=1}^{N}\frac {\delta \lambda_{j}}{\delta u}
=\binom {c_{k_0}}{ b_{k_0}}-\binom {
<\Psi_1,\Phi_2>}{ <\Psi_2,\Phi_1>}=0.\tag 3.8c$$
For $k_0=1$, we have
$$\binom {c_{1}}{ b_{1}}=\binom {r}{q}=\binom {
<\Psi_1,\Phi_2>}{ <\Psi_2,\Phi_1>}=0.\tag 3.9$$
By substituting (3.9) into (3.8a) and (3.8b), the first binary $x$-constrained
 flow becomes a
$x$-FDHS [25]
$$ \Phi_{1x}=\frac {\p F_1}{\p \Psi_1},\quad \Phi_{2x}=\frac {\p F_1}{\p 
\Psi_2},\quad
 \Psi_{1x}=-\frac {\p F_1}{\p \Phi_1},\quad
 \Psi_{2x}=-\frac {\p F_1}{\p \Phi_2},\tag 3.10$$
with the Hamiltonian
$$F_1=<\La\Psi_2, \Phi_2>-<\La\Psi_1, \Phi_1>
+<\Psi_2, \Phi_1><\Psi_1, \Phi_2>.$$\par
Under the constraint (3.9) and the FDHS (3.10), the binary $t_2$-constrained
 flow obtained from  (3.3) with $V^{(2)}$ given by (3.5) and its adjoint 
equation for $N$ distinct real number $\la_j$ can also be written as a 
$t_2$-FDHS
$$ \Phi_{1,t_2}=\frac {\p F_2}{\p \Psi_1},\quad \Phi_{2,t_2}=\frac 
{\p F_2}{\p \Psi_2},\quad
 \Psi_{1,t_2}=-\frac {\p F_2}{\p \Phi_1},\quad
 \Psi_{2,t_2}=-\frac {\p F_2}{\p \Phi_2},\tag 3.11$$
with the Hamiltonian
$$F_2=<\La^2\Psi_2, \Phi_2>-<\La^2\Psi_1, \Phi_1>
+<\Psi_2, \Phi_1><\La \Psi_1, \Phi_2>$$
$$+<\La \Psi_2, \Phi_1><\Psi_1, \Phi_2>
-\frac 12(<\Psi_2, \Phi_2>-<\Psi_1, \Phi_1>)<\Psi_2, \Phi_1><\Psi_1, 
\Phi_2>.$$\par
The Lax representation for the FDHSs (3.10) and (3.11) which can also be
 deduced from the adjoint representation (2.2) and (2.8) are presented by 
(2.18) with the entries of the
 Lax matrix $M$ given by [29]
$$A(\la)=-1+\frac{1}{2}\sum_{j=1}^{N}\frac{\psi_{1j}\phi_{1j}-\psi_{2j}
\phi_{2j}}
{\la-\la_{j}},\tag 3.12a$$
$$B(\la)=\sum_{j=1}^{N}\frac{\psi_{2j}\phi_{1j}}{\la-\la_{j}},\qquad
C(\la)=\sum_{j=1}^{N}\frac{\psi_{1j}\phi_{2j}}{\la-\la_{j}}. \tag 3.12b$$
A straightforward calculation yields
$$A^2(\la)+B(\la)C(\la)\equiv P(\la)=1+
\sum_{j=1}^{N}[\frac{P_{j}}{\la-\la_{j}}+\frac{P^2_{N+j}}{(\la-\la_{j})^2}],
 \tag 3.13$$
where $P_j, j=1,...,2N$, are $2N$ independent integrals of motion for the 
FDHSs (3.10) and (3.11)
$$P_j=\psi_{2j}\phi_{2j}-\psi_{1j}\phi_{1j}$$
$$+\frac 12\sum_{k\neq j}\frac{1}{\la_j-\la_{k}}[(\psi_{1j}\phi_{1j}
-\psi_{2j}\phi_{2j})(\psi_{1k}\phi_{1k}
-\psi_{2k}\phi_{2k})
+4\psi_{1j}\phi_{2j}\psi_{2k}\phi_{1k}],\qquad j=1,...,N\tag 3.14a$$
$$P_{N+j}=\frac 12(\psi_{1j}\phi_{1j}+\psi_{2j}\phi_{2j}),\qquad j=1,...,N.
\tag 3.14b$$
It is easy to verify that
$$F_1=\sum_{j=1}^{N}(\la_jP_{j}+P^2_{N+j})-\frac 14(\sum_{j=1}^{N}P_{j})^2,
 \tag 3.15a$$
$$F_2=\sum_{j=1}^{N}(\la_j^2P_{j}+2\la_jP^2_{N+j})
-\frac 12(\sum_{j=1}^{N}P_{j})
\sum_{j=1}^{N}(\la_jP_{j}+P^2_{N+j})+\frac 18(\sum_{j=1}^{N}P_{j})^3. 
\tag 3.15b$$\par
With respect to the standard Poisson bracket it is found that
$$\{A(\la), A(\mu)\}=\{B(\la), B(\mu)\}=\{C(\la), C(\mu)\}=0,\tag 3.16a$$
$$\{A(\la), B(\mu)\}=\frac 1{\la-\mu}[B(\mu)-B(\la)],\tag 3.16b$$
$$\{A(\la), C(\mu)\}=\frac 1{\la-\mu}[C(\la)-C(\mu)],\tag 3.16c$$
$$\{B(\la), C(\mu)\}=\frac 2{\la-\mu}[A(\mu)-A(\la)].\tag 3.16d$$
Then $\{A^2(\la)+B(\la)C(\la),  A^2(\mu)+B(\mu)C(\mu)\}=0$  implies that 
$P_j, j=1,...,2N,$ are in involution. The AKNS equation (3.6) is factorized 
by the $x$-FDIHS (3.10) and the $t_2$-FDIHS (3.11), namely,
if $(\Psi_1, \Psi_2, \Phi_1,
\Phi_2)$ satisfies the FDIHSs (3.10) and (3.11) simultaneously, 
then $(q, r)$ given by (3.9) solves the AKNS equation (3.6). In general,
 the factorization of the n-th AKNS equations (3.4) will be presented 
in the end of section 3.2.\par

\ \par
\subhead {3.2 The separation of variables for the AKNS equations }
\endsubhead\par
\ \par
In contrast with the $B(\la)$ in the Lax matrix $M$ for the constrained KdV
 flows, the $B(\la)$ given by (3.12b) has only $N-1$ zeros,  one has to 
define the canonical variables $u_k, v_k, k=1,...,2N,$ in a little different 
way. The commutator relations (3.16) and the generating function of integrals
 of motion (3.13) enable us   to introduce
$u_1,...,u_N$ by means of $B(\la)$ in the following way:
$$B(\la)=\sum_{j=1}^{N}\frac{\psi_{2j}\phi_{1j}}
{\la-\la_{j}}=e^{u_N}\frac {R(\la)}{K(\la)},\tag 3.17a$$
where
$$R(\la)=\prod_{k=1}^{N-1}(\la-u_{k}),\qquad
K(\la)=\prod_{k=1}^{N}(\la-\la_{k}),$$
and $v_1,...,v_N$ by
$$v_k= A(u_k), \quad k=1,...,N-1,\quad
v_N=\frac 12(<\Psi_{1}, \Phi_{1}>-<\Psi_{2}, \Phi_{2}>).\tag 3.17b$$
The equation (3.17a) yields
$$u_N=ln<\Psi_{2}, \Phi_{1}>.\tag 3.17c$$
Then it is easy to verify that
$$\{u_N, B(\mu)\}=\{v_N, A(\mu)\}=0,\qquad \{v_N, u_N\}=1,\tag 3.18a$$
$$\{u_N, A(\mu)\}=-\frac {B(\mu)}{<\Psi_{2}, \Phi_{1}>},\qquad
\{v_N, B(\mu)\}=B(\mu).\tag 3.18b$$
As we will show later, the commutator relations (3.16) and (3.18) guarantee 
that $u_1,...,u_N,$ $ v_1,..., v_N$ satisfy the canonical conditions (1.1).\par
We now need  to construct two functions $\widetilde B(\la), \widetilde A(\la)$
 to define $u_{N+1},...,u_{2N}$ by means of $\widetilde B(\la)$
 and $v_{N+1},...,v_{2N}$ by $v_{N+k}=\widetilde A(u_{N+k})$. By 
the exactly same argument as in the previous section,
we construct  $\widetilde A(\la)$ by
$$\widetilde A(\la)=\frac{1}{2}\sum_{j=1}^{N}\frac{\psi_{1j}\phi_{1j}
+\psi_{2j}\phi_{2j}}
{\la-\la_{j}}=\sum_{j=1}^{N}\frac{P_{N+j}}{\la-\la_{j}},\tag 3.19a$$
since the equation (3.19a) enable us  to obtain immediately the separated
 equations (1.2) for $v_{N+k}$ and $u_{N+k}$, and
$\widetilde B(\la)$ by
$$\widetilde B(\la)=\sum_{j=1}^{N}\frac{\phi_{1j}^2}
{\la-\la_{j}}.\tag 3.19b$$\par
Then it is easy to verify that $A(\la), B(\la), \widetilde A(\la), \widetilde
 B(\mu)$ satisfy the commutator relations
$$\{\widetilde B(\la), \widetilde B(\mu)\}=
\{\widetilde B(\la), B(\mu)\}=
\{\widetilde A(\la), \widetilde A(\mu)\}=
\{\widetilde A(\la), B(\mu)\}=
\{\widetilde A(\la), A(\mu)\}=0,\tag 3.20a$$
$$\{\widetilde A(\la), \widetilde B(\mu)\}=\frac 1{\la-\mu}[\widetilde 
B(\mu)-\widetilde B(\la)],\tag 3.20b$$
$$\{A(\la), \widetilde B(\mu)\}=\frac 1{\la-\mu}[\widetilde B(\mu)-\widetilde
 B(\la)].\tag 3.20c$$
The relation (3.20c) doesn't fit the requirement for the canonical conditions
 (1.1). According to 
(3.20) and (3.16) we can replace $A(\la)$ by $\overline A(\la)$
$$\overline A(\la)\equiv A(\la)-\widetilde A(\la)=
-1-\sum_{j=1}^{N}\frac{\psi_{2j}\phi_{2j}}
{\la-\la_{j}},\tag 3.21$$
namely, we redefine $v_1,...,v_N$ by
$$v_k= \overline A(u_k)=A(u_k)-\widetilde A(u_k)=-1-\sum_{j=1}^{N}
\frac{\psi_{2j}\phi_{2j}}
{u_k-\la_{j}}, \quad k=1,...,N-1,\tag 3.22a$$
$$v_N=-<\Psi_{2}, \Phi_{2}>.\tag 3.22b$$\par
We now define $u_{N+1},...,u_{2N}$ by $\widetilde B(\la)$ as follows:
$$\widetilde B(\la)=\sum_{j=1}^{N}\frac{\phi_{1j}^2}
{\la-\la_{j}}=e^{u_{2N}}\frac {\overline R(\la)}{K(\la)},\qquad
\overline R(\la)=\prod_{k=1}^{N-1}(\la-u_{N+k}),\tag 3.23a$$
and  $v_{N+1},...,v_{2N}$ by
$$v_{N+k}=\widetilde A(u_{N+k})=
\frac{1}{2}\sum_{j=1}^{N}\frac{\psi_{1j}\phi_{1j}+\psi_{2j}\phi_{2j}}
{u_{N+k}-\la_{j}}, \quad k=1,...,N-1,\tag 3.23b$$
$$v_{2N}=\frac 12(<\Psi_{1}, \Phi_{1}>+<\Psi_{2}, \Phi_{2}>).\tag 3.23c$$
The equation (3.23a) leads to
$$u_{2N}=ln<\Phi_{1}, \Phi_{1}>.\tag 3.23d$$\par
Then a straightforward calculation shows that $\overline B(\la)=B(\la), 
\overline A(\la), \widetilde B(\la), \widetilde A(\la)$ satisfy the  
commutator relations (2.33) and 
$$\{u_N, \overline B(\mu)\}=\{u_{N}, \widetilde B(\mu)\}=\{u_{N}, 
\widetilde A(\mu)\}=0,\quad
\{u_N, \overline A(\mu)\}=-\frac {\overline B(\mu)}{<\Psi_{2}, \Phi_{1}>},
\tag 3.24a$$
$$\{v_N, \overline A(\mu)\}=\{v_{N}, \widetilde B(\mu)\}=
\{v_{N}, \widetilde A(\mu)\}=0, \quad \{v_N, \overline B(\mu)\}=\overline
 B(\mu),\tag 3.24b$$
$$\{u_{2N}, \overline B(\mu)\}=\{u_{2N}, \overline A(\mu)\}=\{u_{2N}, 
\widetilde B(\mu)\}=0,
 \quad \{u_{2N}, \widetilde A(\mu)\}=-\frac {\widetilde B(\mu)}{<\Psi_{1}, 
\Phi_{1}>},\tag 3.24c$$
$$\{v_{2N}, \overline B(\mu)\}=\{v_{2N}, \overline A(\mu)\}=\{v_{2N},
 \widetilde A(\mu)\}
=0,\quad
\{v_{2N}, \widetilde B(\mu)\}= \widetilde B(\mu),\tag 3.24d$$
$$\{v_N, u_N\}=1,\quad\quad \{v_{2N}, u_{2N}\}=1,\tag 3.24e$$
$$\{u_{2N}, u_N\}=\{u_{2N}, v_N\}=\{v_{2N}, u_N\}=\{v_{2N}, v_N\}=0.
\tag 3.24f$$\par
We have the following proposition.
\proclaim {Proposition 3} 
Assume that  $\la_j, \phi_{ij}, \psi_{ij} \in\text {\bf R}, i=1,2, j=1,...,N$.
Introduce the separated variables
$u_{1},...,u_{2N}$ by the $\overline B(\la)$ and $\widetilde B(\la)$:
$$\overline B(\la)=B(\la)=\sum_{j=1}^{N}\frac{\psi_{2j}\phi_{1j}}
{\la-\la_{j}}=e^{u_N}\frac {R(\la)}{K(\la)},\tag 3.25a$$
$$\widetilde B(\la)=\sum_{j=1}^{N}\frac{\phi_{1j}^2}
{\la-\la_{j}}=e^{u_{2N}}\frac {\overline R(\la)}{K(\la)},\tag 3.25b$$
with
$$R(\la)=\prod_{k=1}^{N-1}(\la-u_{k}), \qquad \overline R(\la)
=\prod_{k=1}^{N-1}(\la-u_{N+k}),$$
and  $v_{1},...,v_{2N}$ by
$$v_{k}=\overline A(u_k)=A(u_k)-\widetilde A(u_k)=
-1-\sum_{j=1}^{N}\frac{\psi_{2j}\phi_{2j}}
{u_k-\la_{j}}, \quad k=1,...,N-1,\tag 3.25c$$
$$v_{N}=-<\Psi_{2}, \Phi_{2}>,\tag 3.25d$$
$$v_{N+k}=\widetilde A(u_{N+k})=
\frac{1}{2}\sum_{j=1}^{N}\frac{\psi_{1j}\phi_{1j}+\psi_{2j}\phi_{2j}}
{u_{N+k}-\la_{j}}, \quad k=1,...,N-1,\tag 3.25e$$
$$v_{2N}=\frac 12(<\Psi_{1}, \Phi_{1}>+<\Psi_{2}, \Phi_{2}>).\tag 3.25f$$
If $u_1,...,u_N,$ are single zeros of $\overline B(\la)$, then
$v_1,...,v_{2N}$ and $u_1,...,u_{2N}$ are canonically conjugated, i.e., 
they satisfy (1.1).\par
\endproclaim\par
{\it{Proof.}} By using the exactly same method as in the proof of proposition
 1,  the commutator relations (2.33) guarantee that $u_1,...,u_{N-1},$ $
 v_1,..., v_{N-1}$ satisfy the canonical conditions (1.1). By using the 
similar method, for example, it is found from (3.24) that for $k=1,...,N-1,$ 
we have
$$0=\{u_N, \overline B(u_k)\}=\overline B'(u_k)\{u_N, u_k\}+\{u_N, \overline
 B(\mu)\}|_{\mu=u_k}=\overline B'(u_k)\{u_N, u_k\},$$
$$\{u_N, v_k\}=\{u_N, \overline A(u_k)\}=\overline A'(u_k)\{u_N, u_k\}- 
\frac {\overline B(\mu)}{<\Psi_{2}, \Phi_{1}>}|_{\mu=u_k}=0,$$
which gives rise to $\{u_N, u_k\}=\{u_N, v_k\}=0$ and so on. In this way 
 we complete the proof. \par
It follows from (3.25a) and (3.25b) that
$$ \psi_{2j}\phi_{1j}=e^{u_{N}}\frac{R(\la_j)}{K'(\la_{j})},\qquad
 \phi_{1j}^2=e^{u_{2N}}\frac{\overline R(\la_j)}{K'(\la_{j})},
\qquad j=1,...,N,$$
or
$$ \phi_{1j}=\sqrt{\frac{e^{u_{2N}}\overline R(\la_j)}{K'(\la_{j})}},\qquad
\psi_{2j}=\frac {e^{u_{N}}R(\la_j)}{\sqrt{e^{u_{2N}}\overline R(\la_j)
K'(\la_{j})}},\qquad j=1,...,N.\tag 3.26$$
The equation (3.9) and (3.17c) results
$$q=e^{u_{N}}.\tag 3.27$$\par
We now present the separated equations.
By substituting $u_k$ into (3.13), $u_{N+k}$ into (3.19a) and using (3.25c) 
and (3.25e), one gets the separated equations
$$v_k=A(u_k)-\widetilde A(u_k)=\sqrt {P(u_k)}-\widetilde A(u_k)$$
$$=\sqrt {1+\sum_{j=1}^{N}[\frac{P_{j}}{u_k-\la_{j}}+\frac{P_{N+j}}
{(u_k-\la_{j})^2}]}
  -\sum_{j=1}^{N}\frac{P_{N+j}}{u_k-\la_{j}},\quad k=1,...,N-1, \tag 3.28a$$
$$v_{N+k}=\widetilde A(u_{N+k})
=\sum_{j=1}^{N}\frac{P_{N+j}}{u_{N+k}-\la_{j}}, 
\qquad k=1,...,N-1. \tag 3.28b$$
It is easy to see from (3.14) that
$$<\Psi_{2}, \Phi_{2}>-<\Psi_{1}, \Phi_{1}>=\sum_{i=1}^{N}P_{i},\qquad
<\Psi_{1}, \Phi_{1}>+<\Psi_{2}, \Phi_{2}>=2\sum_{i=1}^{N}P_{N+i},$$
which together with (3.25d) and (3.25f) leads to
$$v_N=-\frac 12\sum_{i=1}^{N}(P_i+2P_{N+i}),\qquad v_{2N}=\sum_{i=1}^{N}
P_{N+i}.\tag 3.28c$$
 Replacing $v_k$ by the partial derivative $\frac {\p S}{\p u_k}$ of the 
generating function $S$ of the canonical transformation and interpreting 
the $P_i$ as integration constants, 
the equations (3.28) may be integrated to give the generating function of
 the canonical transformation
$$S(u_1,...,u_{2N})=\sum_{k=1}^{N-1}[\int^{u_k}(\sqrt {P(\la)}-\widetilde
 A(\la))d\la
+\int^{u_{N+k}}\widetilde A(\la)d\la]$$
$$-\frac 12\sum_{i=1}^{N}(P_i+2P_{N+i}) u_N+\sum_{i=1}^{N}P_{N+i} u_{2N}$$
$$=\sum_{k=1}^{N-1}[\int^{u_k}\sqrt {P(\la)}d\la-\sum_{i=1}^{N}P_{N+i}ln\mid
\frac{u_k-\la_i}{u_{N+k}-\la_i}\mid]$$
$$-\frac 12\sum_{i=1}^{N}(P_i+2P_{N+i}) u_N+\sum_{i=1}^{N}P_{N+i} u_{2N}.
 \tag 3.29$$\par
The linearizing coordinates are then
$$Q_i=\frac {\p S}{\p P_i}=\frac 12\sum_{k=1}^{N-1}\int^{u_k}\frac {1}{(\la-
\la_i)\sqrt { P(\la)}}d\la-\frac 12u_{N}, \quad i=1,...,N, \tag 3.30a$$
$$Q_{N+i}=\frac {\p S}{\p P_{N+i}}
=\sum_{k=1}^{N-1}[\int^{u_k}\frac {P_{N+i}}{(\la-\la_i)^2\sqrt { P(\la)}}d\la
-ln\mid\frac{u_k-\la_i}{u_{N+k}-\la_i}\mid]-u_{N}+u_{2N},$$
$$ \quad i=1,...,N. \tag 3.30b$$\par
By using (3.15a), the linear flow induced by the FDIHS (3.10)
together with the equations (3.30) leads to the Jacobi inversion problem for 
the FDIHS (3.10)
$$\sum_{k=1}^{N-1}\int^{u_k}\frac {1}{(\la-\la_i)\sqrt {P(\la)}}d\la-u_{N}
=\gamma_{i}+(2\la_i-\sum_{k=1}^{N}P_k)x,
\quad i=1,...,N, \tag 3.31a$$
$$\sum_{k=1}^{N-1}[\int^{u_k}\frac {P_{N+i}}{(\la-\la_i)^2\sqrt { P(\la)}}d\la
-ln\mid\frac {u_k-\la_i}{u_{N+k}-\la_i}\mid]-u_{N}+u_{2N}
=\gamma_{N+i}+2P_{N+i}x,$$ $$\quad i=1,...,N. \tag 3.31b$$\par
By using (3.15b), the linear flow induced by the FDIHS (3.11)
and the equations (3.30) result in the Jacobi inversion problem for the FDIHS 
(3.11)
$$\sum_{k=1}^{N-1}\int^{u_k}\frac {1}{(\la-\la_i)\sqrt { P(\la)}}d\la-u_{N}$$
$$=\bar\gamma_i+[2\la_i^2
-\sum_{k=1}^{N}(\la_kP_k+\la_iP_k+P^2_{N+k})
+\frac 34(\sum_{k=1}^{N}P_k)^2]t_2,
\quad i=1,...,N, \tag 3.32a$$
$$\sum_{k=1}^{N-1}[\int^{u_k}\frac {P_{N+i}}{(\la-\la_i)^2\sqrt { P(\la)}}d
\la-ln\frac {u_k-\la_i}{u_{N+k}-\la_i}]-u_{N}+u_{2N}
$$ $$=\bar\gamma_{N+i}+P_{N+i}(4\la_i-\sum_{k=1}^{N}P_k)t_2,\quad i=1,...,N. 
\tag 3.32b$$\par
Then, since the AKNS equations (3.6) are factorized by the FDIHSs (3.10) and 
(3.11),  combining the equations (3.31) with the equations (3.32) gives rise
 to the 
 Jacobi inversion problem for the AKNS equations (3.6) 
$$\sum_{k=1}^{N-1}\int^{u_k}\frac {1}{(\la-\la_i)\sqrt {P(\la)}}d\la-u_{N}
$$$$=\gamma_{i}+(2\la_i-\sum_{k=1}^{N}P_k)x
+[2\la_i^2
-\sum_{k=1}^{N}(\la_kP_k+\la_iP_k+P^2_{N+k})
+\frac 34(\sum_{k=1}^{N}P_k)^2]t_2,$$
$$\quad i=1,...,N, \tag 3.33a$$
$$\sum_{k=1}^{N}[\int^{u_k}\frac {P_{N+i}}{(\la-\la_i)^2\sqrt { P(\la)}}d\la
-ln\mid\frac {u_k-\la_i}{u_{N+k}-\la_i}\mid]-u_{N}+u_{2N}$$
$$=\gamma_{N+i}+2P_{N+i}x+P_{N+i}(4\la_i-\sum_{k=1}^{N}P_k)t_2,
\quad i=1,...,N. \tag 3.33b$$
If $\phi_{1j}, \psi_{2j}, q$ defined by (3.26) and (3.27) can be solved from 
(3.33)
by using the Jacobi inversion technique, then $\phi_{2j}, \psi_{1j}$ can be 
obtained from the first equation  and the last equation in (3.10) by an 
algebraic calculation, respectively. Finally $q$ and $r=<\Psi_1,\Phi_2>$ 
provides the solution to the AKNS equations (3.6).\par
In general, in order to obtain the Jacobi inversion problem for the n-th 
AKNS equations (3.4),
we set
$$A^2(\la)+B(\la)C(\la)=
\sum_{k=0}^{\infty}\widetilde F_k\la^{-k}, \tag 3.34$$
where $\widetilde F_k, k=1,2,...,$ are also integrals of motion for both the 
FDHS (3.10) and the $t_n$-binary constrained flow. Comparing (3.34) with 
(3.13), one gets
$$\widetilde F_0=1,\quad \widetilde F_k=\sum_{j=1}^{N}[\la_j^{k-1}P_j+(k-1)
\la_j^{k-2}P_{N+j}^2],\quad k=1,2,.... \tag 3.35$$
The n-th AKNS equations (3.4) are factorized by the $x$-FDIHS (3.10) and 
the $t_n$-FDIHS with the Hamiltonian $F_n$ given by [25]
$$F_n=2\sum_{m=0}^{n}(-\frac 12)^{m}\frac {\alpha_m}{m+1}\sum_{l_1+...+l_{m+1}
=n+1}\widetilde F_{l_1}...\widetilde F_{l_{m+1}}, \tag 3.36$$
where $l_1\geq 1,...,l_{m+1}\geq 1,$ and $\alpha_m$ are given by (2.25c).
In the same way, we have
the following proposition.
\proclaim {Proposition 4} The Jacobi inversion problem for the n-th AKNS 
equations (3.4) is of the form
$$\sum_{k=1}^{N-1}\int^{u_k}\frac {1}{(\la-\la_i)\sqrt {P(\la)}}d\la-u_{N}
=\gamma_{i}+(2\la_i-\sum_{k=1}^{N}P_k)x$$
$$+2t_n\sum_{m=0}^{n}(-\frac 12)^m\alpha_m\sum_{l_1+...+l_{m+1}=n+1}
\la_i^{l_{m+1}-1}\widetilde F_{l_1}...\widetilde F_{l_{m}}, 
\quad i=1,...,N, \tag 3.37a$$
$$\sum_{k=1}^{N}[\int^{u_k}\frac {P_{N+i}}{(\la-\la_i)^2\sqrt { P(\la)}}d\la
-ln\mid\frac {u_k-\la_i}{u_{N+k}-\la_i}\mid]-u_{N}+u_{2N}
=\gamma_{N+i}+2P_{N+i}x$$
$$+4t_n\sum_{m=0}^{n}(-\frac 12)^m\alpha_m\sum_{l_1+...+l_{m+1}
=n+1}(l_{m+1}-1)\la_i^{l_{m+1}-2}P_{N+i}\widetilde F_{l_1}...\widetilde 
F_{l_{m}}, 
\quad i=1,...,N, \tag 3.37b$$
where $l_1\geq 1,...,l_{m+1}\geq 1,$ and $
\widetilde F_{l_1},...\widetilde F_{l_{m}},$ are given by (3.35). 
\endproclaim\par

\ \par
\subhead {4. The separation of variables for the Kaup-Newell equations }
\endsubhead\par

\subhead {4.1 Binary constrained flows of the Kaup-Newell hierarchy}
\endsubhead\par
\ \par
For the Kaup-Newell spectral problem [38]
$$\phi_x=U(u,\la)\phi,\quad
 U(u, \lambda)
=\left( \matrix -\la^2&q\la\\r\la&\la^2\endmatrix\right),\quad 
\phi=\binom {\phi_{1}}{\phi_{2}},\quad u=\binom {q}{r},\tag 4.1$$
its adjoint representation (2.2) and (2.3) yields
$$a_{0}=1,\quad a_{2}=-\frac 12 qr,\quad b_{1}=-q,\quad c_1=-r,\quad
 b_{3}=\frac 12(q^2r+q_x),\quad c_3=\frac{1}{2}(qr^2-r_x),...,$$
and in general $a_{2k+1}=b_{2k}=c_{2k}=0$
$$\binom{c_{2k+1}}{ b_{2k+1}}=L \binom{c_{2k-1}}{ b_{2k-1}}, \qquad 
a_{2k}=\frac 12\p^{-1}(qc_{2k-1,x}+rb_{2k-1,x}),\qquad
 k=1,2,\cdots,\tag 4.2$$
$$L=\frac 12\left( \matrix \p-r\p^{-1}q\p&-r\p^{-1}r\p\\-q\p^{-1}q\p&-
\p-q\p^{-1}r\p\endmatrix\right).$$\par
Take
$$\phi_{t_n}=V^{(n)}(u,\la)\phi,\qquad V^{(n)}(u, \la)=\sum_{i=0}^{n-1}
\left( \matrix a_{2i}\la^{2n-2i}&b_{2i+1}\la^{2n-2i-1}\\c_{2i+1}\la^{2n-2i-1}
&-a_{2i}\la^{2n-2i}\endmatrix\right).\tag 4.3$$
Then the compatibility condition of equations (4.1) and (4.3) gives rise 
to the Kaup-Newell hierarchy
$$u_{t_n}={\binom {q}{r}}_{t_n}
=J\binom {c_{2n-1}}{ b_{2n-1}}
=J\frac {\delta H_{2n-2}}{\delta u},\qquad n=1,2,\hdots, \tag 4.4$$
where the Hamiltonian $H_n$ and the Hamiltonian operator $J$ are given by
$$J=\left( \matrix 0&\p\\\p&0\endmatrix\right),\qquad
H_{2n}=\frac{4a_{2n+2}-rc_{2n+1}-q b_{2n+1}}{2n},\qquad
\binom {c_{2n+1}}{b_{2n+1}}
=\frac {\delta H_{2n}}{\delta u}.$$\par
For $n=2$ we have
$$\phi_{t_2}=V^{(2)}(u,\la)\phi,\qquad
V^{(2)}=\left( \matrix \la^4-\frac 12qr\la^2&-q\la^3+\frac{1}{2}(q^2r+q_x)
\la\\-r\la^3+\frac{1}{2}(qr^2-r_x)\la&
-\la^4+\frac 12qr\la^2\endmatrix\right),\tag 4.5$$
and the coupled derivative nonlinear Schr$\ddot{\text o}$dinger (CDNS)
 equations obtained from the equation (4.4) for $n=2$ read
$$q_{t_2}=\frac 12q_{xx}+\frac 12(q^2r)_x, \qquad r_{t_2}=-\frac 12r_{xx}
+\frac 12(r^2q)_x. \tag 4.6$$\par 
The adjoint Kaup-Newell spectral problem is the equation (2.11) with $U$ 
given by (4.1).
We have [25]
$$\frac {\delta\la}{\delta u}=\binom{\frac {\delta\la}{\delta q}}
{\frac {\delta\la}{\delta r}}=\beta Tr[\left(\matrix \phi_1\psi_1
& \phi_1\psi_2\\ \phi_2\psi_1&\phi_2\psi_2
\endmatrix\right)\frac {\p U(u,\la)}{\p u}]=\beta\binom {\la\psi_1\phi_2}
{\la\psi_2\phi_1}.\tag 4.7$$
\par
The binary $x$-constrained flows of the Kaup-Newell hierarchy (4.4) are 
defined by 
$$ \Phi_{1,x}=-\La^2\Phi_{1}+q\La\Phi_{2},\qquad\Phi_{2,x}=r\La\Phi_{1}
+\La^2\Phi_{2}
,\tag 4.8a$$
$$ \Psi_{1,x}=\La^2\Psi_{1}-r\La\Psi_{2},\qquad\Psi_{2,x}=-q\La\Psi_{1}
-\La^2\Psi_{2},\tag 4.8b$$
$$\frac {\delta H_{k_0}}{\delta u}-
\beta^{-1}\sum_{j=1}^{N}\frac {\delta \lambda_{j}}{\delta u}
=\binom {c_{2k_0+1}}{ b_{2k_0+1}}-\frac 12\binom {
<\La\Psi_1,\Phi_2>}{ <\La\Psi_2,\Phi_1>}=0.\tag 4.8c$$
For $k_0=1$, we have
$$\binom {c_{1}}{ b_{1}}=-\binom {r}{q}=\frac 12\binom {
<\La\Psi_1,\Phi_2>}{ <\La\Psi_2,\Phi_1>}=0.\tag 4.9$$
By substituting (4.9) into (4.8a) and (4.8b), the first binary
 $x$-constrained flow becomes a
FDHS
$$ \Phi_{1x}=\frac {\p F_1}{\p \Psi_1},\quad \Phi_{2x}=\frac 
{\p F_1}{\p \Psi_2},\quad
 \Psi_{1x}=-\frac {\p F_1}{\p \Phi_1},\quad
 \Psi_{2x}=-\frac {\p F_1}{\p \Phi_2},\tag 4.10$$
with the Hamiltonian
$$F_1=<\La^2\Psi_2, \Phi_2>-<\La^2\Psi_1, \Phi_1>
-\frac 12<\La\Psi_2, \Phi_1><\La\Psi_1, \Phi_2>.$$\par
Under the constraint (4.9) and the FDHS (4.10), the binary $t_2$-constrained 
flow obtained from  (4.3) with $V^{(2)}$ given by (4.5) and its adjoint 
equation for $N$ distinct reral numbers $\la_j$ can also be written as a FDHS
$$ \Phi_{1,t_2}=\frac {\p F_2}{\p \Psi_1},\quad \Phi_{2,t_2}=\frac {\p F_2}
{\p \Psi_2},\quad
 \Psi_{1,t_2}=-\frac {\p F_2}{\p \Phi_1},\quad
 \Psi_{2,t_2}=-\frac {\p F_2}{\p \Phi_2},\tag 4.11$$
with the Hamiltonian
$$F_2=-<\La^4\Psi_2, \Phi_2>+<\La^4\Psi_1, \Phi_1>
+\frac 12<\La\Psi_2, \Phi_1><\La^3 \Psi_1, \Phi_2>$$
$$+\frac 12<\La^3 \Psi_2, \Phi_1><\La\Psi_1, \Phi_2>-\frac 1{32}<\La \Psi_2, 
\Phi_1>^2<\La\Psi_1, \Phi_2>^2$$
$$+\frac 18(<\La^2\Psi_2, \Phi_2>-<\La^2\Psi_1, \Phi_1>)<\La\Psi_2, \Phi_1>
<\La\Psi_1, \Phi_2>.$$\par
The Lax representation for the FDHSs (4.10) and (4.11) are presented by 
(2.18) with the entries of the
 Lax matrix $M$ given by
$$A(\la)=1+\frac{1}{4}\sum_{j=1}^{N}\frac{\la_j^2(\psi_{1j}\phi_{1j}
-\psi_{2j}\phi_{2j})}
{\la^2-\la^2_{j}},\tag 4.12a$$
$$B(\la)=\frac 12\la\sum_{j=1}^{N}\frac{\la_j\psi_{2j}\phi_{1j}}{\la^2
-\la^2_{j}},\qquad
C(\la)=\frac 12\la\sum_{j=1}^{N}\frac{\la_j\psi_{1j}\phi_{2j}}{\la^2
-\la^2_{j}}. \tag 4.12b$$
A straightforward calculation yields
$$A^2(\la)+B(\la)C(\la)\equiv P(\la)=1+
\sum_{j=1}^{N}[\frac{P_{j}}{\la^2-\la^2_{j}}+\frac{\la_j^4P^2_{N+j}}
{(\la^2-\la^2_{j})^2}], \tag 4.13$$
where $P_j, j=1,...,2N,$ are $2N$ independent integrals of motion for 
the FDIHSs (4.10) and (4.11)
$$P_j=-\frac 12\la_j^2(\psi_{2j}\phi_{2j}-\psi_{1j}\phi_{1j})+\frac 18
<\La\Psi_2, \Phi_1>\la_j\psi_{1j}\phi_{2j}+\frac 18<\La\Psi_1, \Phi_2>
\la_j\psi_{2j}\phi_{1j}$$
$$+\frac 18\sum_{k\neq j}\frac{1}{\la^2_j-\la^2_{k}}[\la^2_j\la^2_{k}
(\psi_{1j}\phi_{1j}-\psi_{2j}\phi_{2j})(\psi_{1k}\phi_{1k}
-\psi_{2k}\phi_{2k})
+2\la_j\la_{k}(\la^2_j+\la^2_{k})\psi_{1j}\phi_{2j}\psi_{2k}\phi_{1k}],$$ 
$$\qquad j=1,...,N\tag 4.14a$$
$$P_{N+j}=\frac 14(\psi_{1j}\phi_{1j}+\psi_{2j}\phi_{2j}),\qquad j=1,...,N.
\tag 4.14b$$
It is easy to varify that
$$F_1=-2\sum_{j=1}^{N}P_{j},\qquad
F_2=2\sum_{j=1}^{N}(\la_j^2P_{j}+\la_j^4P^2_{N+j})-\frac 12
(\sum_{j=1}^{N}P_{j})^2
, \tag 4.15a$$\par
$$<\Psi_2, \Phi_2>+<\Psi_1, \Phi_1>=4\sum_{j=1}^{N}P_{N+j}.\tag 4.15b$$
By inserting $\la=0$, (4.13) leads to
$$1+\frac 14(<\Psi_2, \Phi_2>-<\Psi_1, \Phi_1>)=\sqrt {P(0)}
=\sqrt{1+\sum_{j=1}^{N}[-P_{j}\la^{-2}_{j}+P^2_{N+j}]}.\tag 4.16$$\par
With respect to the standard Poisson bracket it is found that
$$\{A(\la), A(\mu)\}=\{B(\la), B(\mu)\}=\{C(\la), C(\mu)\}=0,\tag 4.17a$$
$$\{A(\la), B(\mu)\}=\frac {\mu}{2(\la^2-\mu^2)}[\mu B(\mu)-\la B(\la)],
\tag 4.17b$$
$$\{A(\la), C(\mu)\}=\frac {\mu}{2(\la^2-\mu^2)}[\la C(\la)-\mu C(\mu)],
\tag 4.17c$$
$$\{B(\la), C(\mu)\}=\frac {\la\mu}{\la^2-\mu^2}[A(\mu)-A(\la)].\tag 4.17d$$
Then $\{A^2(\la)+B(\la)C(\la),  A^2(\mu)+B(\mu)C(\mu)\}=0$  implies that 
$P_j, j=1,...,2N,$ are in involution. The CDNS equations (4.6) are 
factorized by the $x$-FDIHS (4.10) and the $t_2$-FDIHS (4.11), namely,
if $(\Psi_1, \Psi_2, \Phi_1,
\Phi_2)$ satisfies the FDIHSs (4.10) and (4.11) simultaneously, then 
$(q, r)$ given by (4.9) solves the CDNS equations (4.6). The 
factorization of the n-th Kaup-Newell ewuations (4.4) will be 
presented in the end of section 4.2.\par
\ \par
\subhead {4.2 The separation of variables for the Kaup-Newell 
equations }\endsubhead\par
\ \par
Since the commutator relations (4.17) are quite different from (2.23)
 and (3.16), we have to  modify a little bit of the method presented 
in sections 2 and 3. Let us denote $\widetilde \la=\la^2,\widetilde 
\la_j=\la_j^2$. The entries of the Lax matrix $M$ given by (4.12) can
 be rewritten as
$$A(\widetilde\la)=1+\frac 14(<\Psi_2, \Phi_2>-<\Psi_1, \Phi_1>)+\frac 12
 \widetilde\la A_1(\widetilde\la),\quad B(\widetilde\la)
=\frac 12\sqrt{\widetilde\la}\overline B(\widetilde\la),\tag 4.18a$$
where
$$A_1(\widetilde\la)=\frac{1}{2}\sum_{j=1}^{N}\frac{\psi_{1j}\phi_{1j}
-\psi_{2j}\phi_{2j}}{\widetilde\la-\widetilde\la_{j}},\quad
\overline B(\widetilde\la)=\sum_{j=1}^{N}\frac{\sqrt{\widetilde\la_j}
\psi_{2j}\phi_{1j}}{\widetilde\la-\widetilde\la_{j}}.\tag 4.18b$$
It is easy to see that
$$\{A_1(\widetilde\la), A_1(\widetilde\mu)\}=\{\overline B(\widetilde\la),
 \overline B(\widetilde\mu)\}=0,\tag 4.19a$$
$$\{A_1(\widetilde\la), \overline B(\widetilde\mu)\}=\frac 1{\widetilde
\la-\widetilde\mu}[\overline B(\widetilde\mu)-\overline B(\widetilde\la)].
\tag 4.19b$$
It follows from (4.16) and (4.18a) that
$$A(\widetilde\la)
=\sqrt{1+\sum_{j=1}^{N}[-P_{j}\widetilde\la^{-1}_{j}+P^2_{N+j}]}+\frac 12 
\widetilde\la A_1(\widetilde\la).\tag 4.19c$$
The commutator relations (4.19) and the generating function of integrals 
of motion (4.13) enable us to introduce
$u_1,...,u_N$ in the following way:
$$\overline B(\widetilde\la)=\sum_{j=1}^{N}\frac{\sqrt{\widetilde\la_j}
\psi_{2j}\phi_{1j}}{\widetilde\la-\widetilde\la_{j}}
=e^{u_N}\frac {R(\widetilde\la)}{K(\widetilde\la)},\tag 4.20$$
where
$$R(\widetilde\la)=\prod_{k=1}^{N-1}(\widetilde\la-u_{k}),\qquad
K(\widetilde\la)=\prod_{k=1}^{N}(\widetilde\la-\widetilde\la_{k}),$$
and $v_1,...,v_N$ by
$A_1(\widetilde \la).$\par 
By the exactly same argument as in sections 2 and 3, we
 construct  $\widetilde A(\widetilde\la)$ and $\widetilde B(\widetilde\la)$ 
by
$$\widetilde A(\widetilde\la)=\frac{1}{2}\sum_{j=1}^{N}\frac{\psi_{1j}
\phi_{1j}+\psi_{2j}\phi_{2j}}
{\widetilde\la-\widetilde\la_{j}}=2\sum_{j=1}^{N}\frac{P_{N+j}}{\widetilde
\la-\widetilde\la_{j}},\tag 4.21$$
$$\widetilde B(\widetilde\la)=\sum_{j=1}^{N}\frac{\phi_{1j}^2}
{\widetilde\la-\widetilde\la_{j}},\tag 4.22$$
 and, for the same reason, we have to replace $A_1(\widetilde\la)$ by 
$\overline A(\widetilde\la)$
$$\overline A(\widetilde\la)\equiv A_1(\widetilde\la)-\widetilde 
A(\widetilde\la)=
-\sum_{j=1}^{N}\frac{\psi_{2j}\phi_{2j}}
{\widetilde\la-\widetilde\la_{j}}.\tag 4.23$$
Then we have the following proposition.
\proclaim {Proposition 5} 
Assume that  $\la_j, \phi_{ij}, \psi_{ij} \in\text {\bf R}, i=1,2, j=1,...,N$.
Introduce the separated variables
$u_{1},...,u_{2N}$ by the $\overline B(\widetilde\la)$ and $\widetilde 
B(\widetilde\la)$:
$$\overline B(\widetilde\la)=\sum_{j=1}^{N}\frac{\sqrt{\widetilde \la_j}
\psi_{2j}\phi_{1j}}
{\widetilde\la-\widetilde\la_{j}}=e^{u_N}\frac {R(\widetilde\la)}
{K(\widetilde\la)},\tag 4.24a$$
$$\widetilde B(\widetilde\la)=\sum_{j=1}^{N}\frac{\phi_{1j}^2}
{\widetilde\la-\widetilde\la_{j}}=e^{u_{2N}}\frac {\overline 
R(\widetilde\la)}{K(\widetilde\la)},\tag 4.24b$$
with
$$R(\widetilde\la)=\prod_{k=1}^{N-1}(\widetilde\la-u_{k}), \qquad 
\overline R(\widetilde\la)=\prod_{k=1}^{N-1}(\widetilde\la-u_{N+k}),$$
and  $v_{1},...,v_{2N}$ by
$$v_{k}=\overline A(u_k)=A_1(u_k)-\widetilde A(u_k)=
-\sum_{j=1}^{N}\frac{\psi_{2j}\phi_{2j}}
{u_k-\widetilde\la_{j}}, \quad k=1,...,N-1,\tag 4.24c$$
$$v_{N}=-<\Psi_{2}, \Phi_{2}>,\tag 4.24d$$
$$v_{N+k}=\widetilde A(u_{N+k})=
\frac{1}{2}\sum_{j=1}^{N}\frac{\psi_{1j}\phi_{1j}+\psi_{2j}\phi_{2j}}
{u_{N+k}-\widetilde\la_{j}}, \quad k=1,...,N-1,\tag 4.24e$$
$$v_{2N}=\frac 12(<\Psi_{1}, \Phi_{1}>+<\Psi_{2}, \Phi_{2}>).\tag 4.24f$$
If $u_1,...,u_N,$ are single zeros of $\overline B(\la)$, then
$v_1,...,v_{2N}$ and $u_1,...,u_{2N}$ are canonically conjugated, i.e., they 
 satisfy (1.1).\par
\endproclaim\par
{\it{Proof.}} It follows from (4.24a) and (4.24b) that
$$u_N=ln<\La\Psi_{2}, \Phi_{1}>,\tag 4.25$$
$$u_{2N}=ln<\Psi_{1}, \Phi_{1}>.\tag 4.26$$
By a straightforward calculation, it is found that $\overline
 B(\widetilde\la), \overline A(\widetilde\la), \widetilde B(\widetilde\la), 
\widetilde A(\widetilde\la)$ satisfy the  commutator relations (2.33) with 
$\la, \mu$ replaced by $\widetilde\la, \widetilde\mu$, as well as the 
following commutator relations
$$\{u_N, \overline B(\mu)\}=\{u_{N}, \widetilde B(\mu)\}=\{u_{N},
 \widetilde A(\mu)\}=0,\quad
\{u_N, \overline A(\mu)\}=-\frac {\overline B(\mu)}{<\La\Psi_{2}, 
\Phi_{1}>},\tag 4.27a$$
$$\{v_N, \overline A(\mu)\}=\{v_{N}, \widetilde B(\mu)\}=
\{v_{N}, \widetilde A(\mu)\}=0, \quad \{v_N, \overline B(\mu)\}=
\overline B(\mu),\tag 4.27b$$
$$\{u_{2N}, \overline B(\mu)\}=\{u_{2N}, \overline A(\mu)\}=\{u_{2N}, 
\widetilde B(\mu)\}=0,
 \quad \{u_{2N}, \widetilde A(\mu)\}=-\frac {\widetilde B(\mu)}{<\Psi_{1}, 
\Phi_{1}>},\tag 4.27c$$
$$\{v_{2N}, \overline B(\mu)\}=\{v_{2N}, \overline A(\mu)\}=\{v_{2N}, 
\widetilde A(\mu)\}
=0,\quad
\{v_{2N}, \widetilde B(\mu)\}= \widetilde B(\mu),\tag 4.27d$$
$$\{v_N, u_N\}=1,\quad\quad \{v_{2N}, u_{2N}\}=1,\tag 4.27e$$
$$\{u_{2N}, u_N\}=\{u_{2N}, v_N\}=\{v_{2N}, u_N\}=\{v_{2N}, v_N\}=0.
\tag 4.27f$$\par
Then in the exactly same way as for the propositin 1 and 3, we can 
complete the proof.\par
It follows from (4.24a) and (4.24b) that
$$ \la_j\psi_{2j}\phi_{1j}=e^{u_{N}}\frac{R(\la_j^2)}{K'(\la_{j}^2)},\qquad
 \phi_{1j}^2=e^{u_{2N}}\frac{\overline R(\la_j^2)}{K'(\la_{j}^2)},\qquad 
j=1,...,N,\tag 4.28$$
or
$$ \phi_{1j}=\sqrt{\frac{e^{u_{2N}}\overline R(\la_j^2)}{K'(\la_{j}^2)}},
\qquad
\psi_{2j}=\frac {e^{u_{N}}R(\la_j^2)}{\la_j\sqrt{e^{u_{2N}}\overline 
R(\la_j^2)K'(\la_{j}^2)}},\qquad j=1,...,N.\tag 4.29$$
The equations (4.9) and (4.25) result
$$q=-\frac 12e^{u_{N}}.\tag 4.30$$\par
We now present the separated equations.
By substituting $u_k$ into (4.13), $u_{N+k}$ into (4.21) and using (4.19c), 
(4.24c) and (4.24e), one gets the separated equations
$$v_k=A_1(u_k)-\widetilde A(u_k)=\frac 2{u_k}[\sqrt {\widetilde P(u_k)}
-\sqrt {P(0)}]
  -2\sum_{j=1}^{N}\frac{P_{N+j}}{u_k-\la_{j}^2},\quad k=1,...,N-1,
 \tag 4.31a$$
$$v_{N+k}=\widetilde A(u_{N+k})
=2\sum_{j=1}^{N}\frac{P_{N+j}}{u_{N+k}-\la_{j}^2}, 
\qquad k=1,...,N-1, \tag 4.31b$$
where $P(0)$ are given by (4.16) and
$$\widetilde P(\widetilde\la)=
1+\sum_{j=1}^{N}[\frac{P_{j}}{\widetilde\la-\la^2_{j}}+\frac{\la_j^4
P^2_{N+j}}{(\widetilde\la-\la^2_{j})^2}].$$
It follows from (4.15b), (4.16), (4.24d) and (4.24f) that
$$v_N=2-2\sqrt{P(0)}-2\sum_{i=1}^{N}P_{N+i},\quad \qquad v_{2N}=
2\sum_{i=1}^{N}P_{N+i}.\tag 4.31c$$
 Replacing $v_k$ by the partial derivative $\frac {\p S}{\p u_k}$ of the 
generating function $S$ of the canonical transformation and interpreting
 the $P_i$ as integration constants, 
the equations (4.31) may be integrated to give the generating function of
 the canonical transformation
$$S(u_1,...,u_{2N})=\sum_{k=1}^{N-1}[\int^{u_k}(\frac 2{\widetilde\la}\sqrt
 {\widetilde P(\widetilde\la)}
-\frac 2{\widetilde\la}\sqrt {P(0)}-\widetilde A(\widetilde\la))d\widetilde
\la
+\int^{u_{N+k}}\widetilde A(\widetilde\la)d\widetilde\la]$$
$$+(2-2\sqrt{P(0)}-2\sum_{i=1}^{N}P_{N+i})u_N+2\sum_{i=1}^{N}P_{N+i} u_{2N}$$
$$=\sum_{k=1}^{N-1}[\int^{u_k}\frac 2{\widetilde\la}\sqrt {\widetilde 
P(\widetilde\la)}d\widetilde\la-2\sqrt{P(0)}ln|u_k|-2\sum_{i=1}^{N}
P_{N+i}ln\mid\frac{u_k-\la_i^2}{u_{N+k}-\la_i^2}\mid]$$
$$+(2-2\sqrt{P(0)}-2\sum_{i=1}^{N}P_{N+i})u_N+2\sum_{i=1}^{N}P_{N+i} u_{2N}.
 \tag 4.32$$\par
The linearizing coordinates are then
$$Q_i=\frac {\p S}{\p P_i}=\sum_{k=1}^{N-1}[\int^{u_k}\frac {1}{\widetilde
\la(\widetilde\la
-\la_i^2)\sqrt {\widetilde P(\widetilde\la)}}d\widetilde\la+\frac 1{\la_i^2
\sqrt{P(0)}}ln|u_k|]
+\frac 1{\la_i^2\sqrt{P(0)}}u_N,$$
$$\qquad \quad i=1,...,N, \tag 4.33a$$
$$Q_{N+i}=\frac {\p S}{\p P_{N+i}}
=\sum_{k=1}^{N-1}[\int^{u_k}\frac {2\la_i^4P_{N+i}}{\widetilde\la(\widetilde
\la
-\la_i^2)^2\sqrt {\widetilde P(\widetilde\la)}}d\widetilde\la-\frac
 {2P_{N+i}}{\sqrt{P(0)}}ln|u_k|]
-2ln\mid\frac{u_k-\la_i^2}{u_{N+k}-\la_i^2}\mid]$$
$$-2(\frac {P_{N+i}}{\sqrt{P(0)}}+1)u_{N}+2u_{2N}, \quad \qquad i=1,...,N. 
\tag 4.33b$$\par
By using (4.15a), the linear flow induced by (4.10) together 
with the equation (4.33) leads to the Jacobi inversion problem for the FDIHS
 (4.10)
$$\sum_{k=1}^{N-1}[\int^{u_k}\frac {1}{\widetilde\la(\widetilde\la
-\la_i^2)\sqrt {\widetilde P(\widetilde\la)}}d\widetilde\la+\frac 1{\la_i^2
\sqrt{P(0)}}ln|u_k|]
+\frac 1{\la_i^2\sqrt{P(0)}}u_N=\gamma_i-2x,$$
$$ \qquad\quad i=1,...,N, \tag 4.34a$$
$$\sum_{k=1}^{N-1}[\int^{u_k}\frac {2\la_i^4P_{N+i}}{\widetilde\la
(\widetilde\la
-\la_i^2)^2\sqrt {\widetilde P(\widetilde\la)}}d\widetilde\la-\frac 
{2P_{N+i}}{\sqrt{P(0)}}ln|u_k|]
-2ln\mid\frac{u_k-\la_i^2}{u_{N+k}-\la_i^2}\mid]$$
$$-2(\frac {P_{N+i}}{\sqrt{P(0)}}+1)u_{N}+2u_{2N}=\gamma_{N+i}, \qquad
 \quad i=1,...,N. \tag 4.34b$$\par
By using (4.15a), the linear flow induced by (4.11) 
and the equation (4.34) yield the Jacobi inversion problem for the FDIHS 
(4.11)
$$\sum_{k=1}^{N-1}[\int^{u_k}\frac {1}{\widetilde\la(\widetilde\la
-\la_i^2)\sqrt {\widetilde P(\widetilde\la)}}d\widetilde\la+\frac 1
{\la_i^2\sqrt{P(0)}}ln|u_k|]
+\frac 1{\la_i^2\sqrt{P(0)}}u_N=\bar\gamma_i+(2\la_i^2-\sum_{k=1}^{N}P_k)t_2, 
$$
$$\qquad\quad i=1,...,N, \tag 4.35a$$
$$\sum_{k=1}^{N-1}[\int^{u_k}\frac {2\la_i^4P_{N+i}}{\widetilde\la
(\widetilde\la
-\la_i^2)^2\sqrt {\widetilde P(\widetilde\la)}}d\widetilde\la-\frac 
{2P_{N+i}}{\sqrt{P(0)}}ln|u_k|]
-2ln\mid\frac{u_k-\la_i^2}{u_{N+k}-\la_i^2}\mid]$$
$$-2(\frac {P_{N+i}}{\sqrt{P(0)}}+1)u_{N}+2u_{2N}=\bar\gamma_{N+i}+4\la_i^4
P_{N+i}t_2, \quad \qquad i=1,...,N. \tag 4.35b$$\par
Finally, since the CDNS equations (4.6) are factorized by the FDIHS (4.10) 
and (4.11),  combining the equation (4.34) with the equation (4.35) gives 
rise to the 
Jacobi inversion problem for the CDNS equations (4.6) 
$$\sum_{k=1}^{N-1}[\int^{u_k}\frac {1}{\widetilde\la(\widetilde\la
-\la_i^2)\sqrt {\widetilde P(\widetilde\la)}}d\widetilde\la+\frac 1{\la_i^2
\sqrt{P(0)}}ln|u_k|]
+\frac 1{\la_i^2\sqrt{P(0)}}u_N$$
$$=\gamma_i-2x+(2\la_i^2-\sum_{k=1}^{N}P_k)t_2,
\qquad \quad i=1,...,N, \tag 4.36a$$
$$\sum_{k=1}^{N-1}[\int^{u_k}\frac {2\la_i^4P_{N+i}}{\widetilde\la
(\widetilde\la
-\la_i^2)^2\sqrt {\widetilde P(\widetilde\la)}}d\widetilde\la-\frac 
{2P_{N+i}}{\sqrt{P(0)}}ln|u_k|]
-2ln\mid\frac{u_k-\la_i^2}{u_{N+k}-\la_i^2}\mid]$$
$$-2(\frac {P_{N+i}}{\sqrt{P(0)}}+1)u_{N}+2u_{2N}=\gamma_{N+i}+4\la_i^4
P_{N+i}t_2, \quad \quad i=1,...,N. \tag 4.36b$$
If $\phi_{1j}, \psi_{2j}, q$ defined by (4.29) and (4.30) can be solved 
from (4.36)
by using the Jacobi inversion technique, then $\phi_{2j}, \psi_{1j}$ can 
be obtained from the first equation  and the last equation in (4.10),
 respectively. Finally $q$ and $r=-<\La\Psi_1,\Phi_2>$ provides the 
solution to the CDNS equations (4.6).\par
In general, the above procedure can be applied to the whole Kaup-Newell
 hierarchy (4.4). 
Set
$$A^2(\la)+B(\la)C(\la)=
\sum_{k=0}^{\infty}\widetilde F_k\la^{-2k}, \tag 4.37a$$
where $\widetilde F_k, k=1,2,...,$ are also integrals of motion for both 
the $x$-FDHSs (4.10) and the $t_n$-binary constrained flows (2.16). 
Comparing (4.37a) with (4.13), one gets
$$\widetilde F_0=1, \quad \widetilde F_k=\sum_{j=1}^{N}[\la_j^{2k-2}P_j
+(k-1)\la_j^{2k}P_{N+j}^2],\quad k=1,2,.... \tag 4.37b$$
By employing the method in [34,35], the $t_n$-FDIHS obtained from the
 $t_n$-constrained flow  is  of the form
$$ \Phi_{1,t_n}=\frac {\p F_{n}}{\p \Psi_1},\quad \Phi_{2,t_n}=\frac 
{\p F_{n}}{\p \Psi_2},\quad
 \Psi_{1,t_n}=-\frac {\p F_{n}}{\p \Phi_1},\quad
 \Psi_{2,t_n}=-\frac {\p F_{n}}{\p \Phi_2},\tag 4.38a$$
with the Hamiltonian
$$F_{n}=2\sum_{m=0}^{n-1}(-\frac 12)^{m}\frac{\alpha_m}{m+1}
\sum_{l_1+...+l_{m+1}=n}\widetilde F_{l_1}...\widetilde F_{l_{m+1}}, 
\tag 4.38b$$
where $l_1\geq 1,...,l_{m+1}\geq 1,$ and  $\alpha_m$ are given by (2.25c).
Since the n-th Kaup-Newell equations (4.4) is factorized by the $x$-FDIHS 
(4.10) and the $t_n$-FDIHS (4.38). We have the
following proposition.
\proclaim {Proposition 6} The Jacobi inversion problem for the n-th 
Kaup-Newell equations (4.4) is given by
$$\sum_{k=1}^{N-1}[\int^{u_k}\frac {1}{\widetilde\la(\widetilde\la
-\la_i^2)\sqrt {\widetilde P(\widetilde\la)}}d\widetilde\la+\frac 1{\la_i^2
\sqrt{P(0)}}ln|u_k|]
+\frac 1{\la_i^2\sqrt{P(0)}}u_N
=\gamma_i-2x$$
$$+2t_n\sum_{m=0}^{n-1}(-\frac 12)^m\alpha_m\sum_{l_1+...+l_{m+1}=n}
\la_i^{2l_{m+1}-2}\widetilde F_{l_1}...\widetilde F_{l_{m}}, 
\quad i=1,...,N, \tag 4.39a$$
$$\sum_{k=1}^{N-1}[\int^{u_k}\frac {2\la_i^4P_{N+i}}{\widetilde\la
(\widetilde\la
-\la_i^2)^2\sqrt {\widetilde P(\widetilde\la)}}d\widetilde\la-\frac 
{2P_{N+i}}{\sqrt{P(0)}}ln|u_k|]
-2ln\mid\frac{u_k-\la_i^2}{u_{N+k}-\la_i^2}\mid]$$
$$-2(\frac {P_{N+i}}{\sqrt{P(0)}}+1)u_{N}+2u_{2N}=\gamma_{N+i}$$
$$+4t_n\sum_{m=0}^{n-1}(-\frac 12)^m\alpha_m\sum_{l_1+...+l_{m+1}=n}
(l_{m+1}-1)\la_i^{2l_{m+1}}P_{N+i}\widetilde F_{l_1}...\widetilde F_{l_{m}}, 
\quad i=1,...,N, \tag 4.39b$$
where $l_1\geq 1,...,l_{m+1}\geq 1,$ and $\widetilde F_{l_1},...
\widetilde F_{l_{m}},$ are given by (4.37b). 
\endproclaim\par
For example,
the third equations in the Kaup-Newell hierarchy with $n=3$ are of the form
$$q_{t_3}=-\frac 14 q_{xxx}-\frac 38(q^3r^2+2qrq_x)_x,\quad
r_{t_3}=-\frac 14 r_{xxx}-\frac 38(r^3q^2-2qrr_x)_x.\tag 4.40$$
The Kaup-Newell equations (4.40) can be factorized by the $x$-FDIHS 
(4.10) and $t_3$-FDIHS with the Hamiltonian $F_3$ defined by
$$F_3=\sum_{j=1}^{N}(2\la_j^4P_{j}+4\la_j^6P^2_{N+j})-[\sum_{j=1}^{N}
(\la_j^2P_{j}+\la_j^4P^2_{N+j})]\sum_{j=1}^{N}P_{j}
+\frac 14(\sum_{j=1}^{N}P_{j})^3. \tag 4.41$$
The Jacobi inversion problem for the equations (4.40) is given by
$$\sum_{k=1}^{N-1}[\int^{u_k}\frac {1}{\widetilde\la(\widetilde\la
-\la_i^2)\sqrt {\widetilde P(\widetilde\la)}}d\widetilde\la+\frac 1{\la_i^2
\sqrt{P(0)}}ln|u_k|]
+\frac 1{\la_i^2\sqrt{P(0)}}u_N$$
$$=\gamma_i-2x+[2\la_i^4-\sum_{j=1}^{N}(\la_j^2P_{j}+\la_i^2P_{j}+\la_j^4
P^2_{N+j})+\frac 34(\sum_{j=1}^{N}P_j)^2]t_3,
\qquad \quad i=1,...,N, $$
$$\sum_{k=1}^{N-1}[\int^{u_k}\frac {2\la_i^4P_{N+i}}{\widetilde\la
(\widetilde\la
-\la_i^2)^2\sqrt {\widetilde P(\widetilde\la)}}d\widetilde\la-\frac 
{2P_{N+i}}{\sqrt{P(0)}}ln|u_k|]
-2ln\mid\frac{u_k-\la_i^2}{u_{N+k}-\la_i^2}\mid]$$
$$-2(\frac {P_{N+i}}{\sqrt{P(0)}}+1)u_{N}+2u_{2N}=\gamma_{N+i}+[8\la_i^6
P_{N+i}
-2\la_i^4P_{N+j}\sum_{j=1}^{N}P_{j}]t_3, \quad \quad i=1,...,N.$$\par
\ \par
\subhead {5. Concluding remarks}\endsubhead\par
\ \par
The method in [2,3,4,8] is not valid for the separation of variables for 
binary constrained flows of soliton hierarchies. This paper proposed a new
 method to solve this problem. For a certain  kind of integrable models, a
 general approach to the separation of variables was proposed in [2,3,4] by
 taking the poles of the properly normalized Baker-Akhiezer function and the
 corresponding eigenvalues of the Lax operator as separated variables. As
 pointed out in [4], there is no guarantee that the separated variables so 
constructed satisfy the canonical conditions (1.1). The method proposed in 
this paper is to start directly from the canonical conditions (1.1) and the 
requirement for the separated equations (1.2). We introduced $2N$ pairs of 
separated variables by means of four functions $\overline B(\la), \overline
 A(\la), \widetilde B(\la)$ and $\widetilde A(\la)$, which are constructed 
in such way that they satisfy certain commutator relations required by the
 canonical conditions (1.1) and $\overline A(\la)$ and $\widetilde A(\la)$ 
are linked to the generating functions of the integrals of motion for the 
models. This method ensures that the separated variables are canonically
 conjugated. We produced two sets of separated equations directly from two
 generating functions of intergrals of motion. It 
seems that the separated equations are intimately connected with the 
generating functions of intergrals of motion.\par
The finite gap solutions or finite-dimensional quasiperiodic solutions for 
the KdV equation was studied in [36] by means of the stationary equation of 
the KdV hierarchy called the Lax-Novikov equation. By the standard Jacobi 
inversion technique [19], the finite-dimensional quasiperiodic solution 
can be given in an explicit form in terms of the Riemann theta functions
 associated with the invariant spectral curve. The Jacobi inversion problem
 (2.45) for the KdV equation is somewhat different from that in [36] due to 
some additional terms. The Jacobi inversion problems for some binary 
constrained flows require the development of the standard Jacobi inversion
 technique in order to solve them explicitly.\par

\  \par

\subhead {Acknowledgments}\endsubhead\par
This work was supported by  the City University of Hong Kong and the Research
 Grants Council of Hong Kong and the Chinese Basic Research Project 
``Nonlinear Science''.
 One of the authors (Y.B.Zeng) wishes to express his gratitude to
Department of Mathematics of the City University of Hong Kong for warm
 hospitality.
\par
\  \par
 
\subhead {References}\endsubhead \par
\item  {1.} Arnol'd, V.I.: Mathematical methods of classical mechanics,
 2nd edition, New-York: Springer,1994. \par
\item  {2.} Sklyanin, E.K.: Separation of variables in the Gaudin model,
 J. Soviet. Math. 47, 2473-2488 (1989).\par
\item  {3.} Kuznetsov, V.B.: Quadrics on real Riemannian spaces of constant 
curvature: separation of variables and connection with Gaudin magnet, 
J. Math. Phys. 33, 3240-3254 (1992).\par
\item  {4.} Sklyanin, E.K.: Separation of variables, Prog. Theor. Phys.
 Suppl. 118, 35-60 (1995).\par
\item {5.} Babelon, O. and Talon, M.: Separation of variables for the 
classical and quantum Neumann model, Nucl. Phys. B 379, 321-339 (1992).\par
\item  {6.} Eilbeck, J.C., Enol'skii, V.Z., Kuznetsov, V.B. and Tsiganov,
 A.V.: Linear $r$-matrix algebra for classical separable systems, J.
 Phys. A: Math. Gen. 27, 567-578 (1994).\par
\item  {7.} Kalnins, E.G., Kuznetsov, V.B. and Willard Miller, Jr: 
Quadrics on complex Riemannian spaces of constant curvature, separation 
on variables, and the Gaudin magnet, J. Math. Phys. 35, 1710-1731 (1994).\par
\item  {8.} Harnad, J. and Winternitz, P.: Classical and quantum integrable 
systems in $\widetilde {gl}(2)^+$ and separation of variables, Commun. 
Math. Phys. 172, 263-285 (1995).\par
\item  {9.} Kulish, P.P., Rauch-Wojciechowski, S. and Tsiganov, A.V.: 
Stationary problems for equation of the KdV type and dynamical $r$-matrices,
 J. Math. Phys. 37, 3463-3482 (1996).\par
\item  {10.} Zeng, Yunbo: The separability and dynamical $r$-matrix for 
the constrained flows of the Jaulent-Miodek hierarchy, Phys. Lett. A 216,
 26-32 (1996).\par
\item  {11.} Zeng, Yunbo: A farmily of  separable Hamiltonian systems and 
their classical dynamical $r$-matrix, Inverse Problems 12, 1-13 (1996).\par
\item {12.} Zeng, Yunbo: Separation of variables for the constrained flows, J.
 Math. Phys. 38, 321-329 (1997).\par
\item  {13.} Adams, M.R., Harnad, J. and Hurtubise, J.: Darboux coordinates
 and Liouville-Arnold integration in loop algebras,  Commun. Math. Phys. 
155, 385-413 (1993).\par
\item  {14.} Adams, M.R., Harnad, J. and Hurtubise, J.: Liouville generating 
function for isospectral Hamiltonian flow in Loop algebras, in: Integrable 
and superintegrable systems, ed. B.A. Kuperschmidt, Singapore: World 
Scientific, 1990.\par
\item  {15.} Harnad, J. and Wisse, M.A.: Isospectral flow in Loop algebras
 and quasiperiodic solution to the sine Gordon equation, 
J. Math. Phys. 34, 3518-3526 (1993).\par
\item  {16.} Wisse, M.A.: Darboux coordinates and isospectral Hamiltonian 
flow for the massivethirring model, Lett. Math. Phys. 28, 287-294 (1993).\par
\item{17.} Zeng, Yunbo: Using factorization to solve soliton equation, J. 
Phys. A: Math. Gen. 30, 3719-3724 (1997).\par
\item{18.} Zeng, Yunbo: The Jacobi inversion problem for soliton equations,
 J. Phys. Soc. Jpn. 66, 2277-2282 (1997).\par
\item  {19.} Dubrovin, B.A.: Theta functions and nonlinear equations, 
Russian Math. Survey 36, 11-92 (1981). \par
\item {20.} Krichever, I.M. and Novikov, S.P.: Holomophic bundles over 
algebraic curves and nonlinear equations, Russian Math. Surveys 32, 53-79
 (1980).\par
\item {21.} Adler, M. and van Moerbeke, P.: Completely integrable systems, 
Euclidean Lie algebras and curve, Adv. Math. 38, 267-317 (1980).\par
\item {22.} Adler, M. and van Moerbeke, P.: Linearization of Hamiltonian 
systems, Jacobi variables and representation theory, ibid. 38, 318-379 
(1980).\par
\item{23.} Ragnisco, O. and  Rauch-Wojciechowski, S.: Restricted flows of
 the AKNS hierarchy, Inverse Problems 8, 245-262 (1992).\par
\item{24.} Zeng, Yunbo: The higher-order constraint and integrable systems 
related to Boussinesq equation, Chinese Science Bulletin 37, 1937-1942 (1992).
\par
\item{25.} Ma, W.X. and Strampp, W.: An explicit symmetry constraint for the
 Lax pairs and the adjoint Lax pairs of AKNS systems, Phys. Lett. A 185, 
277-286 (1994).\par
\item{26.} Ma, W.X.: New finite-dimensional integrable systems by symmetry
 constraint of the KdV equations, J. Phys. Soc. Jpn. 64, 1085-1091 (1995).\par
\item{27.} Ma, W.X., Fuchssteiner, B. and Oevel, W.: A $3\times 3$ matrix 
spectral problem for AKNS hierarchy and its binary nonlinearization, 
Physica A 233, 331-354 (1996).\par
\item{28.} Ma, W.X., Ding, Q., Zhang, W.G. and Lu, B.Q.: Binary 
nonlinearization of Lax pairs of Kaup-Newell soliton hierarchy, IL 
Nuovo Cimento B 111, 1135-1149 (1996).\par
\item{29.} Li, Yishen and Ma, W.X.: Binary nonlinearization of AKNS 
spectral problem under higher-order symmetry constraints, to appear 
in Chaos, Solitons and Fractals.\par
\item {30.} Ma, W.X. and Fuchssteiner, B.: Binary nonlinearization of
 Lax pairs,  in: Nonlinear Physics, ed. E. Alfinito, M. Boiti, L. 
Martina and F. Pempinelli, Singapore: World Scientific, 1996, pp217-224.
\item {31.} Zeng, Yunbo and Ma, W.X.: The construction of canonical 
separated variables for binary constrained AKNS flow, preprint.
\item {32.} Newell, A.C.: Solitons in mathematics and physics, 
Philadelphia: SIAM, 1985.
\item{33.} Zeng, Yunbo and Li, Yishen: The deduction of the Lax 
representation for constrained flows from the adjoint representation,
 J. Phys. A: Math. Gen. 26, L273-L278 (1993).\par
\item{34.} Zeng, Yunbo: New factorization of the Kaup-Newell hierarchy,
 Physica D 73, 171-188 (1994).\par
\item{35.} Zeng, Yunbo: An approach to the deduction of the 
finite-dimensional integrability from the infinite-dimensional 
integrability, Phys. Lett. A 160, 541-547 (1991).\par
\item {36.} Novikov, S.P.: A method for solving the periodic problem
 for the KdV equation and its generalizations, in: Solitons, Topics
 in current physics, ed.  R. Bullough and P. Caudrey, New York: 
Springer-Verlag, 1980, pp. 325-338.\par
\item{37.} Ablowitz, M. and Segur, H.: Solitons and the inverse 
scattering transform, Philadelphia: SIAM, 1981.\par
\item {38.} Kaup, D.J. and Newell. A.C.: An exact solution for a 
derivative nonlinear Schr$\ddot{\text o}$dinger equation, J. Math.
 Phys. 19, 798-801 (1978).\par

\bye
\bye

\bye
\bye